\documentclass[11pt]{article}
\usepackage{cite}
\usepackage{amsmath,amsfonts,amssymb}
\usepackage[small,bf,hang]{caption}
\usepackage{slashed}
\usepackage{mathabx,amsmath}
\usepackage{color}

\def\hybrid{
        \topmargin -20pt
        \oddsidemargin 0pt
        \headheight 0pt \headsep 0pt
        \textwidth 6.25in 
        \textheight 9.5in 
        \marginparwidth .875in
        \parskip 5pt plus 1pt \jot = 1.5ex}

\hybrid

 \csname
@addtoreset\endcsname{equation}{section}

\newcommand{\Q}{{\cal Q}}

\def\moth{\mathsurround=0pt}
\newdimen\zo \zo=0pt

\def\tick{\leaders\hrule height 0.5ex depth 0pt \hskip 0.5pt}
\def\upboxfill{$\moth \setbox\zo\hbox{\tick}%
  \hskip 3pt\hbox to 0pt{$\tick$\hss}\hrulefill \hbox to 7.5pt{$\tick$\hss}$}

\def\dtick{\leaders\hrule height .34pt depth 0.5ex \hskip 0.5pt}
\def\downboxfill{$\moth \setbox\zo\hbox{\dtick}%
  \hskip 2pt\hbox to 0pt{$\dtick$\hss}\hrulefill \hbox to 2pt{$\dtick$\hss}$}

\def\bec{\begin{center}}
\def\ec{\end{center}}

\def\D{{\cal D}}

\def\S{{\cal S}}
\def\J{{\cal J}}

\def\dt{{\rm d}t}

\def\cS{{\cal S}}
\def\cdS{{\dot{\cal S}}}

 \def\det{{\rm det\,}}
\def\be{\begin{equation}}
\def\ee{\end{equation}}
\def\bea{\begin{eqnarray}}
\def\eea{\end{eqnarray}}
\def\ba{\begin{array}}
\def\ea{\end{array}}

\begin{document}

\begin{titlepage}
\rightline{}
\rightline{May  2019}
\rightline{HU-EP-19/07}
\rightline{MIT-CTP-5119}   
\begin{center}
\vskip 1.5cm
 {\Large \bf{ Duality Invariant Cosmology to all Orders in $\alpha'$}}
\vskip 1.7cm

{\large\bf {Olaf Hohm${}^{1}$ and Barton Zwiebach${}^2$}}
\vskip 1.6cm

{\it ${}^1$ Institute for Physics, Humboldt University Berlin,\\
 Zum Gro\ss en Windkanal 6, D-12489 Berlin, Germany}\\
\vskip .1cm

\vskip .2cm

{\em $^2$ \hskip -.1truecm Center for Theoretical Physics, \\
Massachusetts Institute of Technology\\
Cambridge, MA 02139, USA \vskip 5pt }

\end{center}

\bigskip\bigskip
\begin{center} 
\textbf{Abstract}

\end{center} 
\begin{quote}  
While the classification of $\alpha'$~corrections of string inspired effective 
theories remains an unsolved problem, we show how  to classify 
duality invariant $\alpha'$~corrections for  
purely time-dependent (cosmological) backgrounds. 
We determine the most 
general duality invariant theory to all orders in~$\alpha'$ 
for the metric, $b$-field, and dilaton.
  The resulting Friedmann equations  are studied 
  when the spatial metric is a time-dependent scale factor times the Euclidean metric  and the $b$-field vanishes. 
  These equations can be integrated perturbatively 
  to any order in~$\alpha'$.  
  We construct non-perturbative solutions and display 
  duality invariant theories featuring 
  string-frame  
  de Sitter vacua.

\end{quote} 
\vfill
\setcounter{footnote}{0}
\end{titlepage}

\tableofcontents


\section{Introduction}

String theory is arguably 
the most promising candidate for a theory of quantum gravity.
As a theory of gravity, 
the prospect for a confrontation between string theory and observation 
seems to be particularly 
promising in the realm of cosmology, where the effects of fundamental physics at very small scales may be 
amplified to very large scales. 
Since the early days of string theory there have been intriguing ideas of how 
its  unique characteristics could play a role in cosmological scenarios \cite{Brandenberger:1988aj,Tseytlin:1991xk,Gasperini:2002bn}, 
ideas that have been revisited and extended recently \cite{Brandenberger:2018xwl,Quintin:2018loc,Brandenberger:2018wbg,Wu:2013sha}. 
Two such features of classical string theory will be central for the present paper: 
(i) the existence 
of dualities that  
for cosmological backgrounds  
send the scale factor $a(t)$ of the universe to $1/a(t)$, 
in sharp contrast to Einstein gravity, and 
(ii) the presence of infinitely many higher-derivative $\alpha'$-corrections.

It is a natural idea that the higher-order corrections of string theory play a role, for instance, in resolving the big-bang singularity. 
The first step is the inclusion of the $\alpha'$ corrections of classical string theory.  
Such corrections have been computed to the first few orders in the 1980's 
\cite{Gross:1986iv,Gross:1986mw,Metsaev:1987zx}, but a complete computation or classification of these corrections  
is presently out of reach. 
In recent years duality-covariant formulations 
of the string spacetime theories (double field theory \cite{Siegel:1993th,Hull:2009mi,Hohm:2010jy,Hohm:2010pp}) have been used to make
progress in describing $\alpha'$ corrections, 
see~\cite{Hohm:2013jaa,Hohm:2014eba, Marques:2015vua,Hohm:2016lim,Lescano:2016grn,Hohm:2016lge,Baron:2017dvb}.
These developments hint at a completely novel kind of geometry where the diffeomorphism invariance of general relativity 
is replaced by a suitably generalized notion of diffeomorphisms, which in turn partly determines the $\alpha'$ corrections.  
These ideas play a key role in the 
`chiral' string theory of \cite{Hohm:2013jaa}, which is the only known gravitational
{\em field theory} that is exactly duality invariant and has infinitely many $\alpha'$
corrections.  
This program, however, has not yet been developed to the point that it 
can deal 
with the set of all 
$\alpha'$ corrections 
for bosonic or heterotic string theory.
(See, however, 
the recent proposal for heterotic string theory \cite{Baron:2018lve}.)

This state of affairs is unsatisfactory since the inclusion of a \textit{finite} number of higher-derivative 
$\alpha'$ corrections is generally insufficient. 
Gravitational theories with a finite number of higher derivatives typically 
display various pathologies that are an artifact of the truncation and not present in the full string theory \cite{Zwiebach:1985uq}.
In this paper we will bypass these difficulties by classifying the higher-derivative 
corrections relevant for cosmology to all orders in $\alpha'$. 
Rather than finding the complete $\alpha'$ corrections in general dimensions and then assuming a purely time-dependent cosmological ansatz, 
we immediately consider the theory reduced to one dimension (cosmic time)  
and determine the complete higher-derivative corrections compatible with duality.
String theories, being duality invariant, must correspond to some particular
points in this theory space.   While we do not know which points those are,
the full space of duality invariant theories is interesting in its own right,
and may exhibit phenomena that are rather general and apply to string  
theory.

Our analysis is based on the result 
by Veneziano and Meissner \cite{Veneziano:1991ek}, 
extended by  Sen~\cite{Sen:1991zi},  
concerning the 
classical field theory of the metric, $b$-field and the
dilaton arising from $D = d+1$ dimensional string theory.  This
field theory 
displays an $O(d,d,\mathbb{R})$ symmetry to all orders in~$\alpha'$ 
provided the fields do not depend on the $d$ spatial coordinates. 
This symmetry, henceforth 
referred to as `duality',  
contains the scale-factor duality $a\rightarrow a^{-1}$.
The work of Meissner~\cite{Meissner:1996sa} implies that in terms of standard fields the duality transformations 
receive $\alpha'$ corrections, but it was shown that to first-order in $\alpha'$ there are new field variables in terms of 
which dualities take the standard form.
We will assume that there are field variables so that duality transformations 
remain unchanged  to all orders in $\alpha'$ (this certainly
 happens in conventional
string field theory variables~\cite{Kugo:1992md}). 
With this assumption we are able to classify completely the duality 
invariant  $\alpha'$ corrections. 
This work amounts to an extension and elaboration of the results
obtained by us in~\cite{Hohm:2015doa}.   
 We  
 use the freedom to perform duality-covariant 
field redefinitions to show that only first-order time derivatives need to be included and that the dilaton does not appear nontrivially, 
thereby arriving at a minimal set of duality invariant higher-derivative terms to all orders. 
We prove that 
at order $\alpha'^{\,k} $ the number of independent invariants, and thus the number of free parameters not determined by $O(d,d,\mathbb{R})$, 
 is given by $p(k+1)-p(k)$, with $p(k)$ the number of 
partitions of the integer $k$.

Let us briefly summarize the core technical results of the first part of the paper.  
The two-derivative spacetime theory for the 
metric $g$, the $b$-field $b$, and the dilaton $\phi$,  
restricted 
to depend only on time,  
is described by the one-dimensional action \cite{Veneziano:1991ek}:
\be
\label{vmsbhy} 
\begin{split}
I_0 \ \ = \ &  \int \dt  \, e^{-\Phi}  \Bigl(  -\dot \Phi^{\,2} - \tfrac{1}{8}  \hbox{tr} \big(\dot{\cal S}^2\big) \,  \Bigr) \,, 
\end{split}
\ee
where $\Phi$ is the $O(d,d,\mathbb{R})$ invariant dilaton, defined by $e^{-\Phi}=\sqrt{\det g}\, e^{-2\phi}$, 
and we introduced  the $O(d,d,\mathbb{R})$ valued matrix 
 \be
\label{etaHdef}
{\cal S} \ \equiv\ \begin{pmatrix}  bg^{-1} & g - b g^{-1} b \\[0.7ex]
 g^{-1} & - g^{-1} b 
\end{pmatrix} \;, 
\ee
in terms of the spatial components $g$ and $b$ of the metric and $b$-field. 
Our classification implies that the most general duality invariant $\alpha'$ corrections 
take the form 
\be
\label{vmsrllct022intro} 
\begin{split}
I \ \equiv \ \     I_0 +  \int {\rm d}t \,\, e^{-\Phi}  &\Bigl(  \,   \alpha'  \, c_{2,0}  \hbox{tr} (\cdS^{4}) \  + \  \alpha'^{\,2}  c_{3,0}  \hbox{tr} (\cdS^{6})  \\[-0.5ex]
&  +  \alpha'^{\,3}  \bigl[c_{4,0} \hbox{tr} \bigl(\cdS^{8})   + c_{4,1} \hbox{tr} (\cdS^{4}) \hbox{tr} (\cdS^{4})  \bigr] \\
&  +  \alpha'^{\,4}  \bigl[c_{5,0} \hbox{tr} \bigl(\cdS^{10})   + c_{5,1} \hbox{tr} (\cdS^{6}) \hbox{tr} (\cdS^{4})  \bigr] \\
&  +  \alpha'^{\,5}  \bigl[c_{6,0} \hbox{tr} \bigl(\cdS^{12})   + c_{6,1} \hbox{tr} (\cdS^{8}) \hbox{tr} (\cdS^{4}) + c_{6,2} (\hbox{tr} (\cdS^{6}))^2 
+ c_{6,3} (\hbox{tr} (\cdS^{4}))^3 \bigr] \\
&   +  \cdots \;, 
\end{split}
\ee  
where only first-order time derivatives of $\S$ need to be included.
Moreover, there are no terms involving $\hbox{tr}\big(\dot{\cal S}^2\big)$. 
The pattern is clear: 
the general term at order $\alpha'^{\, k}$ involves traces
with $2k$ factors of $\cdS$.  Each trace must have an even number of $\cdS$
factors, where the even number cannot be two.  
The $c$'s are a priori undetermined coefficients, duality holding for any
value they may take.  Except for a few of them, their values for
the various string theories are unknown.  
  In establishing the above
result, we made repeated use of  field redefinitions
iteratively in increasing orders of $\alpha'$.  The result is striking in that
only first derivatives of the fields appear in the action.  All duality invariant
terms with more than one time derivative on $\S$
can be redefined
away.  All terms with one or more derivatives of the dilaton can also be redefined
away.  The resulting higher derivative actions are in fact actions with high numbers of fields acted on 
by one derivative each.  This is a major, somehow unexpected, simplification.

In the second part of this paper we investigate this general $\alpha'$-complete theory for  the simplest cosmological 
ansatz, a Friedmann-Lemaitre-Robertson-Walker (FLRW) background whose spatial metric is given by a time-dependent scale factor 
times the Euclidean metric and a vanishing $b$-field. 
The resulting Friedmann equations are determined to all orders in $\alpha'$. 
While there is of course still an infinite number 
of undetermined $c_{k,l}$ parameters  in (\ref{vmsrllct022intro}),  
we can write the equations efficiently in terms of a single function $F(H)$ of the Hubble parameter whose Taylor expansion 
is determined by these coefficients: 
 \be\label{FDEFintro}
  F(H) \ := \ 4d\sum_{k=1}^{\infty}(-\alpha')^{k-1}\, 2^{2k-1}\,c_k\, H^{2k}\;, 
 \ee
where, in terms of (\ref{vmsrllct022intro}),  $c_k :=  c_{k,0} + 2\,d\, c_{k,1}  +  \cdots$.  
The Friedmann equations then take the concise form 
  \be\label{firstorderEQS-vm-bbINTRO} 
  \begin{split}
   \frac{{ d}}{{ d}t}\big(e^{-\Phi} f(H) \big) \ &= \ 0
   \;, \\
  \ddot\Phi + \tfrac{1}{2}H f(H) \ &= \ 0\;, \\
   \dot{\Phi}^2+\, g(H) \ &= \ 0\;, 
  \end{split}
 \ee  
where the functions $f(H)$ and $g(H)$ are determined in terms of 
$F(H)$ as:  
 \be
  f(H) \ := \ F'(H)\;, \qquad g(H) \ := \ HF'(H)-F(H)\;, 
 \ee
where ${}^{\prime}$ denotes differentiation with respect to $H$.  
 
We show that solutions of the lowest-order equations 
given by Mueller in~\cite{Mueller:1989in}   
can be extended, perturbatively, 
to arbitrary order in $\alpha'$. We then turn to the arguably most intriguing implication of our results: the potential existence of 
interesting cosmological solutions that are non-perturbative in $\alpha'$. 
We discuss a general non-perturbative initial-value formulation, and we state conditions on the function $F(H)$ so that the 
theory permits de Sitter vacua (in string frame).  
There are functions satisfying these criteria, and so there are duality invariant theories 
with non-perturbative de Sitter vacua, suggesting that string theory may realize de Sitter in this novel fashion \cite{Essay}.  Note that the Lagrangian for the theory
does not include a cosmological constant term.   
We also note that in pure gravity the possibility of generating inflation
through restricted higher-derivative interactions leading to second-order Friedmann equations 
has been investigated~\cite{Arciniega:2018tnn}. 
Here we do not   
constrain the number of derivatives in the original theory. Rather, duality invariance and
field redefinitions lead to two-derivative equations in the cosmological setting.  

The paper is organized as follows.  In section~\ref{class-of-cos-odd} we review
the two-derivative theory and its equations of motion and then tackle the 
classification problem.  We examine carefully the freedom to perform field
redefinitions, including those of the lapse function $n(t)$, which is usually
set to one by a gauge choice.  In section~\ref{equa-of-mot-and-noe} we derive the equations of motion of the higher-derivative
action restricted to the single trace terms, and then compute the Noether charges
for the global duality symmetry.  Section~\ref{fried-equa-toall} specializes to the 
FLRW metric with zero curvature and derives the Friedmann equations to all orders.
As it turns out, this result is valid even when the action contains the most general multi-trace terms.   The solutions in perturbation theory of $\alpha'$ are determined.  In section~\ref{non-perturb-solu} we consider non-perturbative solutions.  We solve the initial-value problem and show how de Sitter solutions
are possible in the space of duality-invariant theories.    
We conclude this paper with a discussion of possible further generalizations.

\section{Classification of cosmological $O(d,d)$ invariants}\setcounter{equation}{0}
\label{class-of-cos-odd}

Our goal is to classify $O(d,d)$ invariant actions to all order in $\alpha'$, up to field redefinitions. This was partially done in \cite{Hohm:2015doa}, 
whose results will here be completed by allowing for $O(d,d)$ covariant field redefinitions of the lapse function.

\subsection{Review of two-derivative theory}
\label{rev_two_der_the} 
We start with the two-derivative theory for 
metric $g_{\mu\nu}$, antisymmetric $b$-field $b_{\mu\nu}$, and 
the scalar dilaton~$\phi$, 
\be
\label{lavm}
I_0 \ = \   \int d^Dx  \sqrt{-g}\,  e^{-2\phi} \Bigl( R +  4 (\partial \phi)^2 -\frac{1}{12} H^2 \Bigr)\;, 
\ee
where $H_{\mu\nu\rho}  =  3  \partial_{[\mu} b_{\nu\rho]}$. We then drop the dependence on all spatial coordinates,
i.e., we set $\partial_i   =   0$, where  $x^\mu   =  (t, x^i)$, $i  =  1,\ldots, d$, 
and we subject the fields to the ansatz 
\be\label{redansatz}
g_{\mu\nu} \ = \ \begin{pmatrix} -n^2  (t) & 0 \\ 0 & g_{ij} (t) \end{pmatrix}\,, \qquad
b_{\mu\nu} \ = \ \begin{pmatrix} 0 & 0 \\ 0 & b_{ij} (t) \end{pmatrix}\,,   \qquad 
\phi \ = \ \phi (t) \,. 
\ee
The resulting one-dimensional  two-derivative action then takes the $O(d,d)$ invariant 
form \cite{Veneziano:1991ek}\footnote{This result generalizes to the compactification of the $D$-dimensional theory on a $d$-torus to 
$D-d$ dimensions,   
which also exhibits a global $O(d,d)$ symmetry \cite{Sen:1991zi, Maharana:1992my,Hohm:2014sxa}.}   
\be
\label{vmsrllct} 
\begin{split}
I_0 \ \equiv \ \int \dt  \, n \,  \, e^{-\Phi} L_0 \ = \ &  \int \dt \, n \, e^{-\Phi}  {1\over n^2} \Bigl(  -\dot \Phi^{\,2} - \tfrac{1}{8}  \hbox{tr} \big(\dot{\cal S}^2\big) \,  \Bigr) \,, 
\end{split}
\ee
where 
 \be
\label{etaHdef}
{\cal S} \ \equiv \ \eta {\cal H} \ = \ \begin{pmatrix}  bg^{-1} & g - b g^{-1} b \\[0.7ex]
 g^{-1} & - g^{-1} b 
\end{pmatrix} \,, \qquad   
\eta \ = \  \begin{pmatrix} 0 & {1} \\[0.7ex]
 {1} & 0
\end{pmatrix} \;. 
\ee
The field $\S$ is constrained: it satisfies ${\cal S}^2={\bf 1}$. We 
have also defined the $O(d,d)$ invariant dilaton~$\Phi$ via
 \be\label{OddDilaton}
  e^{-\Phi} \ = \ \sqrt{\det g_{ij}} \,e^{-2\phi}\;. 
 \ee
The action is time-reparametrization invariant, with $n(t)$ transforming as a density under time reparameterizations $t\to t-\lambda(t)$:
\be
\delta_\lambda n = \partial_t (\lambda n) \,.
\ee 
A scalar $A$ under this reparameterization transforms as $\delta_\lambda A = \lambda \partial_t A$.  The metric, $b$-field, and the dilaton $\Phi$ are scalars under time
reparameterizations. Defining the covariant time derivative~${\cal D}$, 
\be
{\cal D}  \equiv  {1\over n(t)} {\partial \over \partial t}  \;, 
\ee
one can quickly show that if $A$ is a scalar, so is ${\cal D} A$.  The covariant 
derivative satisfies the usual integration by parts rule
\be
\int  \dt \, n\,   B \, {\cal D} A =  \int \dt \, \partial_t (AB) -  \int  \dt \, n\,   ({\cal D}B)A \,. 
\ee
The action $I_0$ above can now be written in a manifestly reparameterization
invariant form
\be
\label{vmsrllct99} 
\begin{split}
I_0 \ =  \int \dt \, n \, e^{-\Phi}  \Bigl(  -({\cal D}\Phi)^2 - \tfrac{1}{8}  \hbox{tr} \big(({\cal DS})^2\big) \,  \Bigr) \,. 
\end{split}
\ee

The variation of the dilaton yields the dilaton equation of motion $E_\Phi=0$, 
 \be\label{DilatonEQ}
  \delta_{\Phi}I_0 \ = \   
  \int {\rm d}t\, n\,e^{-\Phi}\,\delta\Phi  \, E_\Phi\,, \qquad 
   E_{\Phi} \ \equiv \ 2{\cal D}^2 {\Phi} - ( {\cal D}\Phi)^2
  +\tfrac{1}{8} \hbox{tr} \big( ({\cal DS})^2\big) \, . 
 \ee

Let us explain first in general terms how we vary with respect to $\S$ 
to find field equations.
Consider an arbitrary $\S$ dependent action $I$ of the form
\be
I =  \int {\rm d}t\, n\,e^{-\Phi} {\cal L} (\S)\,,
\ee
where ${\cal L}$ could also depend on other fields.
To vary $I$ one first varies $\S$ {\em as if} it would be an unconstrained
field with an unconstrained variation,  finding an expression of the form: 
 \be
 \label{ISvariation}
\delta_\S I =  \int {\rm d}t\, n\,e^{-\Phi}  \,{\rm tr}  \big(
  \delta {\cal S}\,  F_\S  \big)\,. 
\ee
The equation of motion is {\em not} the vanishing of $F_S$ because
$\delta \S$ is a constrained variation:  
since ${\cal S}^2={\bf 1}$ the variation  $\delta \S$ satisfies
$\delta\S=-\S\delta\S \S$. 
We can write $\delta S$  in terms of an {\em unconstrained} variation $\delta K$ as
 follows:
 \be
 \delta \S = \tfrac{1}{2} \bigl( \delta K - \S \delta K \S \bigr) \,.
 \ee
Note that $\delta K$ cannot be written in terms of $\delta \S$, it has
extra information that drops out in the combination shown in the above
right-hand side.   Substituting $\delta\S$ into the above variation $\delta_\S I$
we now get 
\be\label{calSEQ-vmmm}
  \delta_{\S}I \ = \ \int {\rm d}t\,n \,e^{-\Phi}\,
  {\rm tr}  \big(
  \delta K\, E_{\cal S})\,, \qquad   E_{\cal S} \ = \ \tfrac{1}{2} \bigl( F_{\cal S} - \S F_\S \S \bigr) \,.  
 \ee
The equation of motion is indeed $E_\S = 0$, since $\delta K$ is an unconstrained
variation.  Note that any $E_\S$ 
defining 
an equation of motion for $\S$ satisfies
\be
E_\S \ = \ - \S \, E_\S \, \S \,. 
\ee
There is finally an alternative rewriting of $\delta_\S I$ that can now be
seen to be valid too: 
\be
\label{final-option-deltaS}
\delta_\S I  =  \int {\rm d}t\, n\,e^{-\Phi}  {\rm tr}  \big(
  \delta {\cal S} E_\S  \big)\,. 
\ee
Using the expression for $E_\S$ in terms of $F_S$ and the constraint 
$\delta\S= - \S\delta\S\, \S$ one quickly verifies that $\hbox{tr} \big(
  \delta {\cal S} E_\S  \big)=  \hbox{tr} \big(
  \delta {\cal S} F_\S  \big)$ and comparing with (\ref{ISvariation}) we conclude
  that the above right-hand side is indeed equal to $\delta_{\S} I$.    However, 
  (\ref{final-option-deltaS}) is not the operational way to {\em derive} 
  $E_\S$ by variation.

\medskip  
Applying the above discussion for the case of the action $I_0$ we now find
by direct variation
 \be\label{calSEQ9}
  \delta_{\cal S}I_0 \ = \ \int {\rm d}t\, n\,e^{-\Phi}\,
 {\rm tr}  \big(
  \delta {\cal S} F_\S  \big)\,, \quad \hbox{with} \quad   
F_{\cal S} \equiv  \tfrac{1}{4}  \bigl( {\cal D}^2 {\cal S} - {\cal D}\Phi {\cal DS}\bigr) \, . 
 \ee
The equation of motion is 
now $E_\S =0$ with  $E_{\cal S} = 
\tfrac{1}{2} \bigl( F_{\cal S} - \S F_\S \S \bigr)$.
It is straightforward to simplify the resulting expression for $E_\S$.
For this we note 
from $\S \S ={\bf 1}$ 
that $\S$ and ${\cal DS}$ anti-commute:
\be
\S {\cal DS} =  - ({\cal DS}) \S \,.
\ee
Taking a further derivative of this equation we also find that 
\be
\S (\D^2 \S) \S = - \D^2 \S  - 2 \S (\D\S)^2 \,. 
\ee
It is now a simple matter to show that 
\be
  E_{\cal S} \ \equiv \tfrac{1}{4}  \bigl( \D^2 \S  + \S (\D\S)^2 - (\D \Phi) (\D \S) \bigr) \;. 
 \ee
The lapse equation of motion follows by varying with respect to $n$.  
The variation is written in the form:
\be
\label{lapse-eom}
\delta_n I_0 =  \int  \dt \, n \ e^{-\Phi}  {\delta n\over n} \, E_n \,. 
\ee
This is sensible, because one can readily show that a variation $\delta n$ transforms
as a density and, as a result,  the ratio $\delta n/n$ transforms as a scalar.  The quantity
$E_n$ is then a scalar as well.   To compute this variation it is simplest to write
the action $I_0$ in terms of ordinary time derivatives using (\ref{vmsrllct}): 
\be
\label{vmsrllct987} 
I_0 \ \ = \   \int \dt \,  \, e^{-\Phi}  {1\over n} \Bigl(  -\dot \Phi^{\,2} - \tfrac{1}{8}  \hbox{tr} \big(\dot{\cal S}^2\big) \,  \Bigr) \,.
\ee
It follows that 
\be
\label{vct987} 
\delta_n I_0 \ 
\ = \   \int \dt \, n \, e^{-\Phi}  \Bigl( {\delta n\over n} \Bigr) \Bigl(  (\D\Phi)^{2} + \tfrac{1}{8}  \hbox{tr} \big(  ( \D \S)^2\big) \,  \Bigr) \,. 
\ee
Comparing with (\ref{lapse-eom}) we have therefore found that
\be
E_n \, \equiv \,   (\D\Phi)^{2} + \tfrac{1}{8}  \hbox{tr} \big(  ( \D \S)^2\big) \,. 
\ee

\subsection{$O(d,d)$ invariant $\alpha'$ corrections}

We will discuss a recursive procedure to write the duality invariant action
in a canonical form.  In this we use the two-derivative
part of the action to do field redefinitions that allow us to simplify the
higher-derivative parts of the action.  The four-derivative part of the 
action is of order $\alpha'$.  The terms of order $\alpha'^k$ will have $2k+2$ derivatives.  We call $I_{k}$ the part of the action of order 
$\alpha'^k$, and it is 
a sum of terms, 
\be
I_{k}   =  \sum_{p=1}^{N(k)}   I_{k, p}\,, 
\ee
where each term $I_{k, p}$ is a product of dilaton derivatives
and traces of strings of ${\cal S}$ and its derivatives, in order to be
duality invariant.   There are
only a finite number of terms $N(k)$ that can be constructed with
the requisite number of derivatives.  We have
\be
\label{ikp-form}
 I_{k, p} = \alpha'^k \,  c_{k,p}\ 
\int \dt \, n \,    e^{-\Phi} \prod_{i}  (\D^{j_i} \Phi )^{n_i}  \prod_l  \hbox{tr} 
 \Bigl( \bigl(\D^{k^1_l}  {\cal S} \bigr)^{m^1_l}\cdots  
 \bigl( \D^{k^{q_l}_l} {\cal S} \bigr)^{m^{q_l}_l} \Bigr) \,,
\ee
where $c_{k,p}$ are unit-free constants. 
For each $i$ in the first product, the values of the indices $j_i$ and $n_i$ denote the number
of time derivatives and the power of the term.  Because of the dilaton 
theorem, we take $j_i \geq 1$;  all non-derivative dependence on the 
dilaton is in the exponential prefactor.   For each $l$ in the second product
 we get a trace
factor.  Inside the trace there are $q_l$ factors, each one a  fixed power of 
a multiple time
derivative of ${\cal S}$.   This includes the possibility of some $k_l =0$ which
means no derivatives of ${\cal S}$.  This must be accompanied by $m_l =1$
since ${\cal S}^2=1$.   Implicit above is a sum rule that requires
that the total number of derivatives is $2k+2$. 

We will use a notation that makes clear what are the arguments of 
a function $X$ of fields.  When we write 
\be
X (A, B,  \cdots )  \,,
\ee
we mean that $X$ is strictly a function constructed from $A, B, \cdots$ without 
including time derivatives $\D$  of these functions.  When we write
\be
X (\{ A \} , \{ B \} , \cdots ) \,,
\ee
with braces around the arguments, we mean $X$ is a function of $A, \D A , \D^2 A, \cdots$ as well as $B, \D B, \D^2 B, \cdots$.  Mixed types of arguments are
also possible:  $X (A, \{ B \})$ would depend only on $A$, but possibly on
$B , \D B , \D^2 B, \cdots$.  
In this notation equation (\ref{ikp-form}) would be described as saying that  
\be
I_{k, p}  = \alpha'^{\,k} \ \int \dt \, n \,   e^{-\Phi}  \, X ( \{ \D\Phi\}, \{ {\cal S} \} )  \,,   
\ee
for some choice of $X$.  

\medskip 

Our goal is now to prove that  $I_{(k)}$ can be brought to the form 
 \be
 \label{goal-form}
  I_{k} \ = \ \ \alpha'^{\,k} \int \dt \, n \, e^{-\Phi}X(\D\S )\;. 
 \ee
This means that all we can have is a product of traces of powers of $\D\S$.
The proof will proceed inductively, assuming that all terms up to $I_{(k-1)}$ are of this form 
and then using field redefinitions to show that the action $I_{(k)}$ can be brought to this form
(at the cost of changing the actions $I_{k+1}$, $I_{k+2}$, etc., which will be taken care of 
in the next induction step).  

To this end we use various simple theorems established in \cite{Hohm:2015doa}: 
\be\label{generalREL}
\hbox{tr} (\S) \, = \, 
\hbox{tr} (\D \S) \, = \, \hbox{tr} (\D^2 \S) \, = \,  \cdots \, = \,  0 \,.
\ee
\be
\label{oddpowertrace} 
\hbox{tr} ((\D\S)^{2k+1} ) \ = \ 0 \,,  \hbox{   for} \quad k=0,1, \ldots \,. 
\ee
 \be\label{SRELL}
  {\rm tr}({\cal S} (\D \S)^k ) \ = \ 0\,, \hbox{   for} \quad k=0,1, \ldots \,. 
 \ee

\bigskip

The strategy will be to work at each order of $\alpha'$.   Assume that
we have succeeded in casting the action in the expected form (\ref{goal-form})
for $I_2, \cdots I_{k-1}$.  Then we are faced with bringing $I_k$ to the desired
form.  The action can be written as
\be
\label{act-to-start}
I =  I_0 +  \sum_{p=1}^{k-1} I_p  + \alpha'^{\,k}  \int \dt \, n \, e^{-\Phi}  X ( \{ \D \phi\}, \{ \S\} )  + {\cal O} (\alpha'^{\,k+1})\,.   
\ee 
We will then consider field redefinitions of the form
\be
\Phi \to \Phi +\alpha'^{\,k} \delta \Phi  \,,  \qquad  \S \to \S +\alpha'^{\,k} \delta \S \, .  
\ee
To see how $I$ changes, we need only vary $I_0$, because this will generate
${\cal O} (\alpha'^{\,k} )$ terms that we are interested in.  The variations of
$I_1, \cdots , I_{k-1}$ generate terms of higher order that $\alpha'^{\,k} $ that are
not relevant, but are of the prescribed form (\ref{ikp-form}).  The variations of $I_0$
are determined by the field equations and take the form
\be
\label{deltaI00}
\delta I_0 =  \alpha'^{\,k}  \int \dt \, n \, e^{-\Phi}  \big( \delta \Phi E_\Phi + \hbox{tr} ( \delta K  E_\S)  \big)    \quad  \hbox{with} \quad \delta \S = \tfrac{1}{2} \bigl( \delta K - \S \delta K \S \bigr)   \,.  
\ee
At each step we use these field redefinitions to remove undesirable terms 
from the ${\cal O}(\alpha'^{\,k} )$ part of the action.

We elaborate and systematize 
some of the derivations of \cite{Hohm:2015doa}, carefully 
proving several statements that lead to the desired claim.

\begin{enumerate}

\item
  A factor $\D^2{\Phi}$ in an action can be replaced by a factor $Q_\Phi$ with only first derivatives.

Consider a generic such term $Z_k$ at order $\alpha'^{\,k} $ as part of 
(\ref{act-to-start}): 
 \be\label{I_k}
Z_k \ = \   \alpha'^{\,k} \int  \dt \, n \,e^{-\Phi}\,  
  X ( \{ \D\Phi\}, \{ {\cal S} \} )  
 \,  \D^2{\Phi}\;. 
 \ee
With a dilaton field redefinition 
equation (\ref{deltaI00}) gives us 
 \be\label{DilatonEQ93}
\delta  I_0 \ =   
  \alpha'^{\,k} \int {\rm d}t\, n\,e^{-\Phi}\, 2\delta\Phi  \,\bigl( {\cal D}^2 {\Phi} -\tfrac{1}{2}  ( {\cal D}\Phi)^2
  +\tfrac{1}{16} \hbox{tr} \big(({\cal DS})^2\big) \bigr) \,.
 \ee 
If we choose 
 \be
2\delta\Phi \ = \ - X  ( \{\D \Phi\}, \{ {\cal S} \} )\;, 
 \ee
the shift in $I_0$ cancels $Z_k$ completely, and replaces it by
 \be
  Z'_k \ = \   \alpha'^{\,k} \int {\rm d}t\, n\, 
  e^{-\Phi} X ( \{ \D\Phi\}, \{ {\cal S} \} ) \big(\tfrac{1}{2}(\D\Phi)^2 -\tfrac{1}{16}{\rm tr}\, \big( ( \D\S)^2 \big)\big)\,.     
 \ee
The net effect is that the following replacement is allowed in the ${\cal O}(\alpha'^{\,k} )$
action:
 \be\label{doubledotPHI}
  \D^2 {\Phi} \ \rightarrow \  Q_\Phi \equiv \tfrac{1}{2}(\D{\Phi})^2 - \tfrac{1}{16} \hbox{tr} \big(({\cal DS})^2\big)\;, 
 \ee 
with $Q_\Phi$ having only first derivatives.
 
\item  A factor $\D^2 \S$ in an action can be replaced by a factor 
$Q_\S$ with only first order derivatives.

Consider a generic term $Z_k$ at order $\alpha'^{\,k} $ with a second derivative on $\S$: 
 \be
 \label{I_ksder}
Z_k \ = \   \alpha'^{\,k} \int  \dt \, n \,e^{-\Phi}\,  
  X ( \{ \D\Phi\}, \{ {\cal S} \} )  
 \,  \hbox{tr} \bigl( {\cal G}  \,\D^2\S) 
 \,, 
 \ee
 where ${\cal G}$ is a {\em matrix} of type ${\cal G} ( \{  \S\})$.  
With a field redefinition of $\S$,
equation (\ref{deltaI00}) gives: 
\be\label{calSEQ-red}
  \delta I_0 \ = \  \alpha'^{\,k} \int {\rm d}t\,n \,e^{-\Phi}\,
  {\rm tr}  \big[\, \tfrac{1}{4} \, 
  \delta K\, \bigl( \D^2 \S  + \S (\D\S)^2 - (\D \Phi) (\D S) \bigr)\bigr] \,.  
 \ee
Choosing the matrix $\delta K$ (which determines $\delta \S = \tfrac{1}{2} (\delta K - \S \delta K \S$))  to be
\be
\tfrac{1}{4} \, 
  \delta K \ = \ -  X ( \{ \D\Phi\}, \{ {\cal S} \} )  \, {\cal G}\,,
\ee 
and realizing that $X$ is a scalar that can go out of the trace,  it gives
\be\label{calSEQ-red-98}
  \delta I_0 \ = \ - \alpha'^{\,k} \int {\rm d}t\,n \,e^{-\Phi}\, X ( \{ \D\Phi\}, \{ {\cal S} \} ) \, 
  {\rm tr}  \big[\, 
 {\cal G} \, \bigl( \D^2 \S  + \S (\D\S)^2 - (\D \Phi) (\D S) \bigr)\bigr] \,.  
 \ee
This variation cancels the $Z_k$  term above and replaces it by
 \be
 \label{I_ksder}
Z'_k \ = \   \alpha'^{\,k} \int  \dt \, n \,e^{-\Phi}\,  
  X ( \{ \D\Phi\}, \{ {\cal S} \} )  
 \,  \hbox{tr} \bigl( {\cal G}  ( - \S (\D\S)^2 + (\D \Phi) (\D S))) \,. 
 \ee
The net effect is that the following replacement is allowed in the ${\cal O}(\alpha'^{\,k} )$
action:
 \be\label{doubledotcalS}
  \D^2 \S \ \rightarrow  Q_\S \equiv  \ -{\cal S}(\D\S)^2 + \D{\Phi}\, \D \S\;, 
 \ee
with $Q_\S$ having at most first derivatives. 

\item  Any action can be reduced so that it only has first time derivatives of $\Phi$.

We already know how to eliminate terms with two derivatives on the dilaton.
Suppose we have an action with more than two derivatives of $\Phi$. This can always be written in the form 
 \be
 \label{zkfinds}
  Z_k  \ =  \alpha'^{\,k} \ \int \dt \, n \,  e^{-\Phi} X(\{\D{\Phi}\},\{{\cal S}\})\, 
  \D^{p+2}     
  \Phi\;,    \quad  0 < p \leq 2k \,
 \ee
where the dilaton factor
to the right has more than two derivatives.    
Now, write $\D^{p+2} \Phi =  \D^{p} (\D^2 \Phi)$ and integrate
by parts the $\D^{p}$, finding 
 \be\label{partialintstep}
  Z_k \ = \  \alpha'^{\,k} \ \int \dt \, n \, 
 (-\D)^{p}\big(e^{-\Phi}\, X\big) 
   \D^2\Phi\;, 
 \ee
suppressing the arguments of $X$.  
We can now use property 1  to eliminate the $\D^2 \Phi$, replacing $Z_k$ by 
\be\label{partialintstep}
  Z'_k  =   \alpha'^{\,k}  \int \dt \, n \, 
 (-\D)^{p}\big(e^{-\Phi}\, X\big)\, 
   \big(\tfrac{1}{2}(\D{\Phi})^2 - \tfrac{1}{16}{\rm tr}\, (\D\S)^2 \big)\, .  
 \ee
Now we integrate back the $p$ derivatives 
of the left factor, one by one. At each step this will generate a second-order time derivative of the second factor, 
but these can be reduced to first derivatives by properties 1 and 2. We can write this
formally defining a derivative operator $\bar \D$ that acts on functions
$F$  
of $\S, \D\S,$ and $\D\Phi$:
\be
(\bar\D F)(\S, \D\S, \D\Phi) \ \equiv \ \D \big(F(\S, \D\S, \D\Phi)\big)\Bigl|_{\D^2 \Phi \to Q_\Phi,\, 
\D^2 \S \to Q_\S} \,. 
\ee
The replacements here are those in (\ref{doubledotPHI}) and (\ref{doubledotcalS}).  Note that 
after taking a $\bar \D$ derivative, the result is still a function of only $\S, \D\S,$ and $\D\Phi$.  It follows now that $Z_k$ in (\ref{zkfinds}) is eventually replaced by
\be\label{partialintstep}
  Z''_k  =   \alpha'^{\,k}  \int \dt \, n \,  e^{-\Phi} 
 X(\{\D{\Phi}\},\{{\cal S}\}) \ \bar\D^{p}  Q_\Phi \, .
 \ee
The factor to the right of $X$ contains only first derivatives of $\Phi$ and
at most first derivatives of $\S$.  Because of (\ref{SRELL}), the factor to the right
of $X$ cannot contain
$\S$, it must be built explicitly with $\D\S$ only.  

Should $X$ above still contain  factors with more than two derivatives
of $\Phi$ the procedure is carried through again for each factor. 
 Proceeding in this way we can reduce the higher derivative factors in $\Phi$ 
one by one to first order in time derivatives.   This step has shown that
the most general form of the
action is 
\be\label{partialintstep-so-far}
  I_k  =   \alpha'^{\,k}  \int \dt \, n \,  e^{-\Phi} 
 X (\D{\Phi},\{{\cal S}\}) \, .  
 \ee
This means that, apart from the $e^{-\Phi}$ factor, only first derivatives of the dilaton appear. 

\item  Any action can be reduced so that it only has first time derivatives of ${\cal S}$.

Assume there is a trace factor that has a time derivative of $\S$ of degree
higher than two.   Using the form of the action (\ref{partialintstep-so-far}) we already know to be true,  we write the action separating out such a term: 
\be
\label{partialintstep-so-far-99}
  Z_k  =   \alpha'^{\,k}  \int \dt \, n \,  e^{-\Phi} 
 X(\D {\Phi},\{{\cal S}\})  
 \ \hbox{tr} [ {\cal G} (\{ \S\}) 
 \D^{p+2} \S ]  \, ,  \quad 0< p \leq  2k \,.   
 \ee
Clearly $X$ must contain $2k-p$ derivatives, but we need not include this
in the notation.  Moreover, ${\cal G}$ is a matrix built as products of $\S$
and its derivatives.   Defining the matrix ${\cal F}$ as follows: 
\be
\label{hoiefn}
{\cal F}(\D{\Phi},\{{\cal S}\}) \equiv  e^{-\Phi}\,  X(\D{\Phi},\{{\cal S}\})\, {\cal G} (\{ \S\})\,,
\ee
the action above takes the form 
\be\label{partialintstep-so-far-991}
  Z_k  =   \alpha'^{\,k}  \int \dt \, n \,  
  \, \hbox{tr} [\, {\cal F} \, \D^{p+2} \S ]   \,.   
 \ee
For brevity, we do not indicate the arguments of ${\cal F}$, which
are given in (\ref{hoiefn}). 
In this form integration by parts is straightforward and folds the derivatives
into ${\cal F}$.  We integrate by parts $\D^p$ and then are allowed
to replace $\D^2 \S$ by $Q_\S$, finding that $Z_k$ is replaced by
\be\label{partialin87tstep-so-far-991}
  Z'_k  =   \alpha'^{\,k}  \int \dt \, n \,  
  \, \hbox{tr} [\,(-\D)^p {\cal F} \, Q_\S ]   \,.   
 \ee
The derivatives are then folded back into $Q_\S$, one at a time, at each
step eliminating the second derivatives that are generated.  As explained
in property 3, each $(-\D)$ derivative becomes a $\bar \D$ derivative and
we find that $Z'_k$ becomes  
\be\label{partialin87tstep-so-far-991}
  Z'_k  =   \alpha'^{\,k}  \int \dt \, n \,  
  \, \hbox{tr} [\, {\cal F} \, \bar\D^pQ_\S ]   \,.   
 \ee
Using the definition of ${\cal F}$ we now note that (\ref{partialintstep-so-far-99})
has become
\be
\label{partialintstep-so-vfar-99}
  Z''_k  =   \alpha'^{\,k}  \int \dt \, n \,  e^{-\Phi} 
 X(\D {\Phi},\{{\cal S}\})  
 \ \hbox{tr} [\,  {\cal G} (\{ \S\}) 
 \bar\D^{p} Q_\S ]  \,,   
 \ee
achieving the desired elimination of the higher derivative of $\S$ in terms
of first derivatives of $\Phi$ and up to first derivatives of $\S$.  This procedure
can be iterated until all higher derivatives of $\S$ are eliminated.  At his point
only $\S$ and $\D\S$ can appear and, as explained before, this means that only
$\D\S$ can appear.   We have thus shown that terms of $I_k$, already
put in the form (\ref{partialintstep-so-far}),  can be simplified further to be set in the form
\be\label{partialintsjnbbtep-so-far}
  I_k  =   \alpha'^{\,k}  \int \dt \, n \,  e^{-\Phi} 
 X (\D{\Phi}, \D\S) \, .  
 \ee
 Each term $I_{k,p}$ in such an action is written in the form 
\be\label{partimmcheryalintsjnbbtep-so-far}
  I_{k,p}  =   \alpha'^{\,k}  \int \dt \, n \,  e^{-\Phi} \, (\D\Phi)^p
 X (\D\S) \, .  
 \ee
We will now show that in fact any dilaton derivative or power of a dilaton
derivative can be eliminated! 

\item 
 Any action $I_k$, with $k>1$ is equivalent to one without any appearance of $\D\Phi$.
 
Suppose we have an action $I_k$ with a term
 \be\label{original-vm}
  I_{k,p} \ = \ \int  \dt \, n \, e^{-\Phi}\, (\D\Phi)^p \,  X_l(\D\S)\;. 
 \ee
Here $X_l$ is an arbitrary (generally multi-trace) invariant:
 \be\label{simplemono}
   X_l(\D\S) \ = \ {\rm tr}\bigl[(\D\S)^{l_1}\bigr]\, \cdots {\rm tr}\bigl[(\D\S)^{l_n}\bigr]\;, \quad l = l_1 + \cdots + l_n\,.
 \ee 
In a moment we will need that the time derivative of this function has a simple form.
As we take a derivative we generate $\D^2 \S$ factors within the traces.  Each such
factor can be replaced by 
$Q_\S = -\S (\D\S)^2 + \D \Phi \D \S$, but since traces
with a single $\S$ multiplying $\D\S$ factors vanish ((\ref{SRELL})), the replacement
is effectively $\D^2\S \to \D\Phi \D\S$.  As a result,  
 inside an integral representing a term in $I_k$,
the following replacement is allowed:
 \be
  \D X_l \ \sim \ l\,\D\Phi\,  X_l(\D\S) \,. 
 \ee
We begin by rewriting (\ref{original-vm}) with one factor of $\D\Phi$ written
as a derivative of $e^{-\Phi}$, and then proceed to do integration by parts:
 \be
 \begin{split}
  I_{k,p} \ &= \   -\int  \dt \, n \, \D (e^{-\Phi})  (\D\Phi)^{p-1} X_l 
 \\
  & = \ \int \dt \, n \,  e^{-\Phi}\big((p-1)(\D\Phi)^{p-2} \D^2{\Phi} \, X_l
  +(\D\Phi)^{p-1}\, \D X_l \big)\,.   
 \end{split} 
 \ee
Replacing $\D^2 \Phi$ by $Q_\Phi$ and using the above evaluation of $\D X_l$ we get
 \be
 \begin{split}
  I_{k,p} 
  \ &\sim \   \int \dt \, n \,   
  e^{-\Phi}\big((p-1)(\D\Phi)^{p-2} (\tfrac{1}{2}(\D\Phi)^2-\tfrac{1}{16}{\rm tr}\,
  (\D\S)^2) X_l\  
  + l\, (\D{\Phi})^{p}{X_l}\big)\\
  \ &= \ \int \dt \, n \,  
  e^{-\Phi}\big(\tfrac{1}{2}(p+2l-1)(\D\Phi)^{p}{X_l}
  - \tfrac{1}{16}(p-1)(\D\Phi)^{p-2}\,  {\rm tr}\,( \D\S)^2  X_l \big) \,. 
 \end{split} 
 \ee
The first term is of the same form as the original (\ref{original-vm}), and thus bringing it to the left-hand side we have 
the equivalence 
 \be
 \tfrac{1}{2} (3-p-2l)\int \dt \, n \,  e^{-\Phi}(\D\Phi)^p X_l \ = \ - \tfrac{1}{16}(p-1)
 \int \dt \, n \, e^{-\Phi} (\D\Phi)^{p-2} {\rm tr}\,(\D\S)^2 \,  X_l\,, 
 \ee
and thus in actions we can replace:
 \be
 \label{master-red-dil}
 \int \dt \, n \,  e^{-\Phi}(\D\Phi)^p X_l(\D\S) \ \sim \ \frac{1}{8}\Big(\frac{p-1}{p+2l -3}\Big)\, \int \dt \, n \, e^{-\Phi} (\D\Phi)^{p-2} {\rm tr}\,(\D\S)^2X_l(\D\S)\,.
 \ee
Let us note that the denominator on the right-hand side never vanishes
for any case of interest.  Since the total number of derivatives in the term 
is $p+l$, 
for a term in an action $I_k$,  the denominator  takes the value
\be
 (p+l) + l-3 = 2k + 2 + l -3 = 2k + l -1 \geq 1\,,  \quad \hbox{for} \ \ k \geq 1 \,.
\ee
The formula can be used recursively to reduce the power of $\D\Phi$ in steps
of two, in each step $p$ decreases by two units and $l$ increases by two units
(since ${\rm tr}\,(\D\S)^2X_l$ is an $X_{l+2}$ object), 
and at no stage will the denominator vanish.  If $p= 2m$ is even, 
it follows from (\ref{master-red-dil}) that we have 
\be
 \label{master-red-dil-vm}
 \int \dt \, n \,  e^{-\Phi}(\D\Phi)^{2m}  X_l(\D\S) \ \sim \ C\, \int \dt \, n \, e^{-\Phi} \bigl[{\rm tr}\,(\D\S)^2\bigr]^m\, 
 X_l(\D\S)\,,
 \ee
 where $C$ is a constant that is easily determined. 
 Note also that for $p=1$ equation (\ref{master-red-dil}) tells us
 that 
\be
 \label{p=1-red-dil}
 \int \dt \, n \,  e^{-\Phi}\, \D\Phi\,  X_l(\D\S) \ \sim \ 0\,.
 \ee
For the case $l=1$, where the prefactor in (\ref{master-red-dil}) gives zero divided 
by zero, the left hand side 
vanishes too because $\hbox{tr} (\D\S) =0$.  Since the recursion relation
works in steps of two units, the vanishing of terms with a single $\D\Phi$ 
implies that 
\be
 \label{master-red-dil-cl}
 \int \dt \, n \,  e^{-\Phi}(\D\Phi)^{2m+1}  X_l(\D\S) \ \sim \ 0\,  . 
 \ee
The formula (\ref{master-red-dil}) also applies when $l=0$ and $X_l=1$, in which
case it gives  
 \be
 \label{master-red-jn}
 \int \dt \, n \,  e^{-\Phi}(\D\Phi)^p \ \sim \ \frac{1}{8}\Big(\frac{p-1}{p-3}\Big)\, \int \dt \, n \, e^{-\Phi} (\D\Phi)^{p-2} {\rm tr}\,(\D\S)^2\,.
 \ee
The denominator in the prefactor only vanishes for $p=3$ which is not relevant
for any action $I_k$.   The remarks above apply again.  When $p=2m$ is even, the term
can be reduced to one containing no dilaton derivatives and just $[ \hbox{tr} (\D\S)^2]^m$.  If $p$ is odd, the term vanishes because it does for $p=1$.  

We  have 
shown so far that any term $I_{k,p}$ in the action $I_k$ takes the
form (\ref{partimmcheryalintsjnbbtep-so-far}).  Since we have now learned
that powers of derivatives of the dilaton can be eliminated completely we have shown
that any term in $I_k$ takes the claimed form (\ref{goal-form}):
\be
 \label{goal-form-got-it}
  I_{k} \ = \ \ \alpha'^{\,k}  \int \dt \, n \, e^{-\Phi}X(\D\S )\;. 
 \ee
 While this is a great simplification, further simplification can be
 achieved by using redefinitions of the lapse function $n(t)$.  We explore this next. 
\end{enumerate}

\subsubsection*{Lapse redefinitions} 

We have confirmed  that up to $O(d,d)$ covariant redefinitions of ${\cal S}$ and $\Phi$ one only needs 
to consider (arbitrary powers of) first time derivatives of ${\cal S}$ and no time derivatives of $\Phi$.  
Thus,  the most general action takes the following form: 
 \be
  I \ = \ \int \dt \, n \, e^{-\Phi}\Big(\,  L_0 + \alpha'\, L_1 + (\alpha')^2 L_2+(\alpha')^3 L_3+\cdots \Big)\;, 
 \ee
where the general form of $L_1, L_2,$ and $L_3$ are: 
 \be\label{Ansaetze}
 \begin{split}
  L_1 \ &= \ a_1{\rm tr}(\D\S)^4+a_2 \, [{\rm tr}(\D\S)^2]^2\,, \\
  L_2 \ &= \ b_1{\rm tr}(\D\S)^6+b_2{\rm tr} (\D\S)^4\,{\rm tr} \, (\D\S)^2
+b_3 [{\rm tr} (\D\S)^2]^3\,, \\
 L_3 \ &= \ c_1{\rm tr}(\D\S)^8
+c_2 [{\rm tr} \, (\D\S)^4]^2 
+c_3{\rm tr}\,(\D\S)^6\,{\rm tr} \, (\D\S)^2\\
& \ \ \ +c_4{\rm tr}\,(\D\S)^4\, [{\rm tr} \, (\D\S)^2]^2
+c_5 [{\rm tr} \,(\D\S)^2]^4\,.
 \end{split} 
 \ee

We will now show that using lapse redefinitions, together with further 
dilaton redefinitions,  we can remove any
term in the action $I$ that has a $\hbox{tr} (\D\S)^2$ factor! This
leaves, as 
one 
can see, 
 the first term in $L_1$, the first term in $L_2$, and the first two terms in 
$L_3$.

As before, we work at fixed orders of $\alpha'$ recursively:  at each fixed order
$\alpha'^{\,k} $ we eliminate terms with a $\hbox{tr} (\D\S)^2$ factor by doing
a redefinition of $n$ of order $\alpha'^{\,k} $ that contributes through
the variation of $I_0$.   This variation was given 
in (\ref{vct987}):
\be
\label{vct987-elab} 
\delta_n I_0 \ 
\ = \   \int \dt \, n \, e^{-\Phi}  \Bigl( {\delta n\over n} \Bigr) \Bigl(  (\D\Phi)^{2} + \tfrac{1}{8}  \hbox{tr} ( \D \S)^2 \,  \Bigr)\,.
\ee
Consider now a general term in $I_{k,p}$ with a $\hbox{tr} (\D\S)^2$ factor:
\be
\label{4irfn}
I_{k,p} \ = \ \alpha'^{\,k}   \int \dt \, n \, e^{-\Phi} X_{2k} (\D\S) \, \hbox{tr} (\D\S)^2\,.
\ee
Here $X_{2k} (\D\S)$ denotes a term with $2k$ derivatives built from products of 
traces of powers of $\D\S$.  It does not matter if $X_{2k}$ has additional
factors of $\hbox{tr} (\D\S)^2$, it may.  We will consider a redefinition of the
form 
\be
{\delta n\over n} = \beta  \alpha'^{\,k} \,  X_{2k} (\D\S)\,,
\ee
where $\beta$ is a constant to be determined.  The associated variation of $I_0$ is
therefore 
\be
\label{vct987-elabxx} 
\delta_n I_0 \ 
\ = \  \beta  \alpha'^{\,k}  \int \dt \, n \, e^{-\Phi}  X_{2k} (\D\S) \Bigl(  (\D\Phi)^{2} + \tfrac{1}{8}  \hbox{tr} (\D \S)^2 \,  \Bigr)\,.
\ee
This counts as an additional term in the action.  Note the second term in parenthesis
gives a contribution that could cancel the term in $I_{k,p}$ above.  The first term
also does, if we use identity (\ref{master-red-dil}) which amounts to a further dilaton redefinition.  This identity with $p=2$ and  $l= 2k$ allows us to write: 
\be
\label{vct987-elabxxx} 
\begin{split}
\delta_n I_0 \ 
\ & = \  \beta  \alpha'^{\,k}  \int \dt \, n \, e^{-\Phi}  X_{2k} (\D\S) \Bigl(  {1\over 8(4k-1)} \hbox{tr} (\D\S)^2 + \tfrac{1}{8}  \hbox{tr} (\D \S)^2 \,  \Bigr)\\
\ & = \   {\beta k \over 2(4k-1)} 
 \alpha'^{\,k}  \int \dt \, n \, e^{-\Phi}  X_{2k} (\D\S)  \hbox{tr} (\D \S)^2 \,. 
\end{split}
\ee
One can cancel $I_{k,p}$ in (\ref{4irfn}) by choosing $\beta$
such that $ {\beta k \over 2(4k-1)}=-1$. Since this can
always be done, this completes our proof that any term with a $\hbox{tr} (\D\S)^2$
factor can be set to zero.

Given this result, the most general $\alpha'$ 
corrections up to order $(\alpha')^3$ given in (\ref{Ansaetze}) 
can be simplified to take the form
\be\label{Ansaetze-improved}
 \begin{split}
  L_1 \ &= \ a_1{\rm tr}(\D\S)^4\,, \\
  L_2 \ &= \ b_1{\rm tr}(\D\S)^6\,, \\
 L_3 \ &= \ c_1{\rm tr}(\D\S)^8
+c_2 [{\rm tr} \, (\D\S)^4]^2\,. 
 \end{split} 
 \ee
Thus, refining the classification of \cite{Hohm:2015doa}, we find that 
the number of independent coefficients in the ${\cal O} ((\alpha')^{k})$
action is given by the number of partitions of $k+1$ that do not use 1, 
which in the literature is sometimes denoted by $p(k+1,2)$. 
Using Euler's generating function for the unrestricted number $p(n)$ of partitions of 
the integer $n$, it is easy to prove that this equals 
 \be
  p(k+1,2) \ = \ p(k+1) \ - \ p(k)\;, 
 \ee  
see, e.g., \cite{partitions}. This proves our statement in the introduction.

\section{Equations of motion and Noether charges}
\label{equa-of-mot-and-noe}

In this section we determine the explicit equations of motion for the special case that only single trace invariants
are included, which is, however, sufficient for the cosmological applications in the next section. 
Moreover, we discuss Bianchi identities and the conserved Noether charges corresponding to $O(d,d)$ duality invariance.

\subsection{Equations of motion}
We now compute the equations of motion obtained from the action (\ref{vmsrllct}) upon 
including single trace higher derivative  terms. Upon setting $n=1$ 
this action reads 
\be
\label{vmsrllct022} 
\begin{split}
I \ \equiv \ \ &  \int {\rm d}t \,\, e^{-\Phi}  \Bigl(  -\dot \Phi^{2} +  \sum_{k=1}^{\infty}  \alpha'^{\,k-1}  c_{k}\, \hbox{tr} (\cdS^{2k}) \,  \Bigr) \,,  
\end{split}
\ee  
where we included the lowest order term in the sum, with $c_1  =  -\frac{1}{8}$. 
The higher $c_k$ coefficients are, of course, only partially known (e.g.~$c_2  =  \frac{1}{64}$ for bosonic string theory, $c_2  =  \frac{1}{128}$ for 
heterotic string theory, and $c_2=0$ for type II string theories).

In order to derive the equation of motion for $\S$ we follow the steps
reviewed in section~\ref{rev_two_der_the}. 
The variation with respect to $\S$ 
of (\ref{vmsrllct022}) reads 
\be
\label{varyS-intheaction}
\begin{split}  
\delta_{\S}I 
\ = & \ \sum_{k=1}^{\infty}  \alpha'^{\,k-1}  c_{k} 
\int {\rm d}t\,  e^{-\Phi}  \,  (2k) \, \hbox{tr} \Bigl( {d\delta{\cal S} \over dt}  \cdS^{2k-1}\Bigr)\\
\ = & \ \sum_{k=1}^{\infty}  \alpha'^{\,k-1}  c_{k} \int {\rm d}t\, e^{-\Phi}  \,  (2k) \, \hbox{tr} \Bigl( {d \over dt} \Bigl[ 
\delta {\cal S} \,  \cdS^{2k-1}\Bigr] - \Bigr[ \delta \S  {d\over dt } \cdS^{2k-1} 
\Bigr]\Bigr)\\
\ = & \ \sum_{k=1}^{\infty}  \alpha'^{\,k-1}  c_{k} \int {\rm d}t\, e^{-\Phi}  \,  \, \hbox{tr} \Bigl( 
\delta {\cal S} \,  (2k)  \Bigl[ \dot \Phi \cdS^{2k-1} -  {d \over dt } \cdS^{2k-1}
\Bigr]\Bigr) \\
\ = & \ \int {\rm d}t\,  e^{-\Phi}  \,  \, \hbox{tr} \Bigl( 
\delta {\cal S} \,  \sum_{k=1}^{\infty}  \alpha'^{\,k-1}  c_{k}(2k)  \Bigl[ \dot \Phi \cdS^{2k-1} -  {d \over dt } \cdS^{2k-1}
\Bigr]\Bigr)\;,
\end{split}
\ee
where we discarded total time derivatives.   The above defines  $F_\S$, and
therefore 
$E_\S$ is given~by 
\be
\begin{split}   
  E_\S = \tfrac{1}{2}  
\sum_{k=1}^{\infty}    
\alpha'^{\,k-1}  c_{k} (2k) \Biggl( \ 
\Bigl[ \dot \Phi \cdS^{2k-1} -  {d \over dt } \cdS^{2k-1}
\Bigr] -  \S\Bigl[ \dot \Phi \cdS^{2k-1} -  {d \over dt } \cdS^{2k-1}
\Bigr]\S \Biggr) \;. 
\end{split}
\ee
This form can be simplified by using the anti-commutativity of ${\cal S}$ and $\cdS$: 
\be\label{noncommSTEP}
\begin{split}
E_\S \ = \    
\sum_{k=1}^{\infty}  \alpha'^{\,k-1}  c_{k}\,  k \Bigl( \ 
2 \dot \Phi \cdS^{2k-1} -  {d \over dt } \cdS^{2k-1}
+  \S {d \over dt } \cdS^{2k-1}
\cS \Bigr) \;. 
\end{split}
\ee
This equation can be further simplified as follows: using $\S\ddot{\S} = -\ddot{\S}\S-2\dot{\S}^2$, 
which follows by taking the second derivative of $\S^2={\bf 1}$, 
it is straightforward to verify 
\be
 \S\,\frac{d}{dt}\dot{\S}^{2k-1}\,\S \ = \ -\frac{d}{dt}\dot{\S}^{2k-1} \ - \ 2\S\dot{\S}^{2k}\;. 
\ee
Using this in (\ref{noncommSTEP}), 
$E_\S$ finally reduces to 
\be\label{allORDEReq}
E_\S \ =  \,  -2\sum_{k=1}^{\infty}  \alpha'^{\,k-1}  c_{k}\,  k \Bigl( 
  {d \over dt } \cdS^{2k-1}\ - \dot \Phi \cdS^{2k-1}
+  \cS  \cdS^{2k}
 \Bigr) \;. 
\ee
The equation of motion for $\S$ is $E_\S=0$.

In order to find the equation of motion for $n$ we restore this dependence in
the action (\ref{vmsrllct022}):
\be
\label{vmsrllct0225} 
\begin{split}
I \ \equiv \ \ &  \int {\rm d}t  \, n \,\, e^{-\Phi}  \Bigl(  -(\D\Phi)^2 +  \sum_{k=1}^{\infty}  \alpha'^{\,k-1}  c_{k}\, \hbox{tr} (\D \S)^{2k} \,  \Bigr) \,,  \\ 
= \ \ &  \int {\rm d}t 
\,\, e^{-\Phi}  \Bigl(  -{1\over n} \dot \Phi^2 +  \sum_{k=1}^{\infty}  {\alpha'^{\,k-1}  \over n^{2k-1} }c_{k}\, \hbox{tr} (\dot\S^{2k}) \,  \Bigr)\;. 
\end{split}
\ee  
One quickly finds by variation of $n$ and then back to covariant notation that
\be
\label{vmsrllct02257} 
I \ \equiv \ \   \int {\rm d}t  \, n \,\, e^{-\Phi}  {\delta n\over n}\Bigl(  (\D\Phi)^2 -  \sum_{k=1}^{\infty}  \alpha'^{\,k-1}  (2k-1) c_{k}\, \hbox{tr} ((\D \S)^{2k} )\,  \Bigr) \,.  
\ee 
The object in between big parenthesis is  $E_n$. 
It is straightforward to compute the $\Phi$ variation of (\ref{vmsrllct022}). 
To recapitulate
let us record the general variation.  Recall that 
$E_\S$ can be used directly in the variation (see~(\ref{final-option-deltaS})), 
and we can set $n=1$ in the $E_n$ equation, too. We have 
 \be\label{generalVAR}
 \delta I \ = \ \int {\rm d}t\,n\,e^{-\Phi}\Big(\delta \Phi E_{\Phi}+ {\rm tr}\big(\delta \S E_{\S}\big)+\frac{\delta n}{n} E_n \Big)\;, 
 \ee
with the (off-shell) functions 
 \be\label{offshellfunctions}
 \begin{split}
  E_{\Phi} \ &\equiv \ 2\ddot{\Phi} -\dot{\Phi}^2 - \sum_{k=1}^{\infty}\alpha'^{\,k-1}  c_k \,{\rm tr}\big(\dot{\S}^{2k}\big)\;, \\
  E_{\S} \ &\equiv \ -2\sum_{k=1}^{\infty}\alpha'^{\,k-1} k\, c_k \Big(\frac{{d}}{{d}t}\dot{\S}^{2k-1}-\dot{\Phi}\,\dot{\S}^{2k-1}+\S \dot{\S}^{2k}\Big)\;, \\
  E_n \ &\equiv \ \dot{\Phi}^2-\sum_{k=1}^{\infty}\alpha'^{\,k-1} (2k-1)c_k\,{\rm tr}\big(\dot{\S}^{2k}\big)\;. 
 \end{split}
 \ee
These expressions satisfy a Bianchi identity as a consequence of time reparametrization invariance of $I$ under the field variations:  
 \be
  \delta_{\xi}\S \ = \ \xi\,\dot{\S}\;, \quad \delta_{\xi}\Phi \ = \ \xi\,\dot{\Phi}\;, \quad 
  \delta_{\xi}n \ = \ \partial_t(\xi n)\;. 
 \ee
Specializing (\ref{generalVAR}) to these variations  and integrating by parts we then infer 
 \be
 \begin{split}
  0 \ = \ \delta_{\xi}I 
  \ = \  \int {\rm d}t
  \,e^{-\Phi}\,\xi\Big(\dot{\Phi} E_{\Phi} + {\rm tr}\big(\dot\S E_{\S}\big)-e^{\Phi}\partial_t\big(e^{-\Phi} E_{n}\big)\Big)
  \;,
  \end{split} 
 \ee
where we set $n=1$, which is legal after variation. Since the integral vanishes for arbitrary $\xi$ we infer the 
 \be\label{LapseBianchi}
 \hbox{Bianchi identity:} \qquad 
  \dot{\Phi}\big(E_{\Phi}+E_{n}\big) + {\rm tr}\big(\dot\S E_{\S}\big) \ =  \ \frac{{ d}}{{ d}t}E_n\;. 
 \ee
This identity may also be verified by a direct computation using (\ref{offshellfunctions}). 
It will be instrumental because it implies that it is sufficient to solve the equations  
 \be\label{finalEQS}
 \begin{split}
  E_{\Phi}+E_{n} \  = \ 0 \,,  \quad  
  \quad E_{\S} \ = \ 0\;, 
 \end{split} 
 \ee
and to impose the lapse equation (or Hamiltonian constraint) 
$E_n=0$ only  on initial data.  
The lapse equation
will hold for all times as a consequence of (\ref{LapseBianchi}).  Note that 
\be
E_{\Phi}+E_{n} \ = \ 2\ddot{\Phi}-\sum_{k=1}^{\infty}\alpha'^{\,k-1} (2k)c_k\,{\rm tr}
 \big(\dot{\S}^{2k}\big)\;. 
\ee

\subsection{Noether charges}
We will now compute the Noether charges corresponding to $O(d,d)$ invariance. 
To this end we consider the infinitesimal $O(d,d)$ transformations that leave 
the action invariant and which are given by 
 \be\label{OddS}
  \delta_{\tau}\S \ = \ \tau \S-\S\tau\;, \qquad \delta_{\tau}\Phi \ = \ 0\;, 
 \ee
where $\tau=(\tau^M{}_N)\in\frak{so}(d,d)$, thus satisfying 
 \be
 \label{condition-on-tau}
   \tau\eta+\eta\tau^t  =  0\;. 
 \ee 
In using matrix notation we also have $S= ( S^M{}_{ N}) = \eta {\cal H}$
with $\eta= (\eta^{MN})$ and ${\cal H}= {\cal H}_{MN}$.
With $h = (h^M{}_{N})$ an $O(d,d)$ group element, 
we have $h\eta h^t = \eta$
and the Lie-algebra valued infinitesimal parameter 
$\tau $ arises from $h \simeq  1 + \tau$   
and thus satisfies (\ref{condition-on-tau}).  
The transformation $\delta_\tau \S$ above  
 is the infinitesimal version of the finite transformation $\S\rightarrow \S'=h \S h^{-1}$, again with $h\simeq 1+\tau$. 
Note that $\delta_\tau \S = - \S \, \delta_\tau \S \, \S$ and thus 
the variation 
preserves  $\S^2={\bf 1}$. 
The symmetry constraint on $\S$ (inherited from ${\cal H} = 
{\cal H}^t$)  reads  
 \be\label{anotherSconstr}
  \S\eta -\eta \S^t \ = \ 0\;, 
 \ee
and is also preserved under the variation $\delta_\tau \S$.

We can compute the Noether charges $\Q$ by the usual trick of promoting the symmetry 
parameter to be local, $\tau\rightarrow \tau(t)$, and to compute the variation 
 \be\label{generalOddVAR}
  \delta_{\tau}I \ = \  \int {\rm d}t\,{\rm tr}\big(\dot{\tau}\Q\big) \ = \ -\int {\rm d}t\,{\rm tr}\big({\tau}\dot{\Q}\big)\;. 
 \ee
An explicit computation of the variation of the action $I$ in 
 (\ref{vmsrllct022}) under $\delta_\tau \S$ 
 follows quickly   
 from the first line on (\ref{varyS-intheaction}) and gives 
 \be\label{NoetherCharge}
  \Q \ = \ e^{-\Phi}\sum_{k=1}^{\infty} \alpha'^{\,k-1} 4\,k\, c_k \,\S\, \dot{\S}^{2k-1}\;. 
 \ee
Since on-shell the variation (\ref{generalOddVAR}) must vanish for arbitrary functions $\tau$, we 
conclude that on-shell $\dot\Q=0$. Thus $\Q$ is conserved, which may also be verified by 
a quick computation with (\ref{allORDEReq}), which gives
$\dot \Q = -2 e^{-\Phi} \S E_\S$, and shows that the conservation of $\Q$
is equivalent to the equation of motion of $\S$.  
Moreover, $\Q$ takes values in the Lie algebra $\frak{so}(d,d)$ 
as one quickly sees that 
$\Q\eta +\eta \Q^t  =  0$, using (\ref{anotherSconstr}) and its time derivative.

\section{Friedmann equations to all orders in $\alpha'$} 
\label{fried-equa-toall}

In this section we evaluate the 
equations of motion (\ref{offshellfunctions}) for
the spatial metric set equal to  
 a scale factor times the Euclidean metric 
and with vanishing $b$-field. 
This yields the $\alpha'$ corrected Friedmann equations. 
In the first subsection  
we write  
the $O(d,d)$ covariant fields in terms of conventional cosmological 
variables  
and show that the standard (string) Friedmann equations are recovered to 
zeroth order in $\alpha'$. In the second subsection we give the $\alpha'$ corrected equations. 
In the final subsection we show how to integrate these equations to arbitrary order in $\alpha'$.

\subsection{Review of two-derivative equations}

We now specialize to the Friedmann-Lemaitre-Robertson-Walker (FLRW) metric 
with curvature $k=0$ 
and vanishing $b$-field:  we set  $g_{ij}=a^2(t) \delta_{ij}$ 
and $b_{ij}=0$ in (\ref{etaHdef}). Together with the definition of the $O(d,d)$ invariant dilaton in (\ref{OddDilaton}) 
we then have 
 \be\label{SFRW}
  {\cal S}(t) \ = \ \begin{pmatrix}  0 & a^2(t) \\[0.7ex]
 a^{-2}(t) & 0 
\end{pmatrix}\;, \qquad
e^{-\Phi} \ = \ (a(t))^d e^{-2\phi} \;, 
 \ee 
in terms of the scale factor $a(t)$ and the scalar dilaton $\phi(t)$. 
We next introduce  the Hubble parameter 
\be
H \ \equiv \ \frac{\dot{a}}{a}\;. 
\ee 
The time derivatives of the dilaton $\Phi$ then become:   
 \be
 \label{phiPhi-der}
  \dot{\Phi} \ = \ -d\, H+2\dot{\phi} \;, \qquad 
  \ddot{\Phi} \ = \ -d\,\dot{H}+2\ddot{\phi}\;, 
 \ee 
where 
 \be\label{HDOTT}
  \dot{H} \ = \ -H^2+\frac{\ddot{a}}{a}\;. 
 \ee 

For the following applications it will be convenient to establish some relations for the field $\S(t)$ and its derivatives. 
First, $\S(t)$ is constrained and satisfies $\S^2={\bf 1}$, and is hence a `pseudo-complex' structure. Amusingly, its derivative is proportional to a 
complex structure: 
 \be\label{SDOTTT}
  \dot\S \ = \ 2H\J\;, \qquad \J \ \equiv \  \begin{pmatrix}  0 & a^2(t) \\[0.7ex]
 -a^{-2}(t) & 0 
\end{pmatrix}\;,   \qquad \J^2 \ = \ - {\bf 1} \,.   
 \ee
In turn, the derivative of $\J$ is proportional to $\S$: 
 \be
  \dot\J \ = \ 2H \S\;. 
 \ee 
Thus, the second time derivative of $\S$ can be expressed in terms of $\S$ and $\J$: 
 \be
  \ddot\S \ = \ 2\dot H \J + 4H^2 \S\;. 
 \ee 

Let us now evaluate the two-derivative equations of motion following from the action $I_0$ for the above ansatz, 
with the aim to compare with the standard Friedmann equations that follow directly
from the higher dimensional action for metric and dilaton. 
After setting $n(t)=1$ the equations of motion 
of the two-derivative theory are those in~(\ref{offshellfunctions}), keeping only
$c_1 = -\tfrac{1}{8}$ and setting all $c_{k>1}$ equal to zero:   
\be
\begin{split}
\ddot{\cal S}+{\cal S}\dot{\cal S}^2-  \,\dot{\Phi}\,\dot{\cal S}\ = \ &  \ 0 \,, \qquad ({\cal S})\\
\ -2\ddot{\Phi}+\dot{\Phi}^{\,2}-\tfrac{1}{8}{\rm tr}\,\dot{\cal S}^2 \ = \ & \  0\,, \qquad (\Phi)\\
\dot{\Phi}^2+\tfrac{1}{8}{\rm tr}\,\dot{\cal S}^2 \ = \ & \  0\,. \qquad (n)
\end{split}
\ee 
Inserting the above  expressions for $\S, \Phi$ and their derivatives in terms
of the scale factor, 
these equations become 
 \be
 \begin{split}
  (d-1)\Big(\frac{\dot{a}}{a}\Big)^2 + \frac{\ddot{a}}{a}-2\frac{\dot{a}}{a}\dot{\phi} \ &= \ 0\,, \qquad ({\cal S})\\
  d(d-1)H^2-4\,d\, H\dot{\phi}+4\,\dot{\phi}^2-4\,\ddot{\phi}+2\,d\,\frac{\ddot{a}}{a} \ &= \ 0\,, \qquad (\Phi)\\
  d(d-1) H^2-4\,d\, H\dot{\phi}+4\,\dot{\phi}^2 \ &= \ 0\,. \qquad (n)
 \end{split}
 \ee 
Replacing the second equation by the third minus the second, we find
  \be\label{FriedmannMAN}
 \begin{split}  
  (d-1)\Big(\frac{\dot{a}}{a}\Big)^2 
  + \frac{\ddot{a}}{a}-2\frac{\dot{a}}{a}\dot{\phi} \ &= \ 0\,, \qquad ({\cal S})\\
 -2\,\ddot{\phi}+\,d\,\frac{\ddot{a}}{a} \ &= \ 0\,, \qquad (n)-(\Phi)  \\
  d(d-1) H^2-4\,d\, H\dot{\phi}+4\,\dot{\phi}^2 \ &= \ 0\,. \qquad (n)
 \end{split}
 \ee 
Using linear combinations of the above equations, it is easy to
see that they are 
equivalent 
to the three standard (string) Friedmann equations, 
as for instance given in~\cite{Yang:2005rw}.

\subsection{$\alpha'$ corrected equations}

Let us now turn to the computation of the $\alpha'$ corrected Friedmann equations. 
We will start from the equations  
in~(\ref{offshellfunctions})  
and specialize to the FLRW ansatz with $k=0$ shown in (\ref{SFRW}). 
Although these equations were derived only from the general single-trace action (\ref{vmsrllct022}) 
we will now argue that for the FLRW ansatz, and leaving the coefficients $c_k$ generic, 
this is in fact the most general action. 

Specifically, we will show that the inclusion of any multi-trace 
invariants in the action merely renormalizes the coefficients $c_k$. To this end note that 
with $\dot\S = 2 H \J$  
and $\J^2=-{\bf 1}$ we can immediately evaluate a generic term of order 
$\alpha'^{\,k-1} $ in the Lagrangian: 
 \be\label{singletraceEV}
  {\cal L} \; \ \propto \; \ c_k\,{\rm tr}\big(\dot{\S}^{2k}\big) \ = \ (-1)^{k}\,2^{2k+1}\,c_k\, d\, H^{2k}(t)\;. 
 \ee
Now, let us compare this with, say, a double trace term with the same order of $\alpha'$: 
 \be\label{doubletraceinv}
  {\cal L} \; \ \propto \; c_{k,l}\,{\rm tr}\big(\dot{\S}^{2(k-l)}\big)\,{\rm tr}\big(\dot{\S}^{2l}\big) \ = \ 
  c_{k,l}(-1)^k 2^{2k+1}\,2\,d^2\, H^{2k}(t)\;, 
 \ee
where $c_{k,l}$ are new coefficients. Comparing with (\ref{singletraceEV}) we see that they coincide, 
except for the overall normalization. Thus, the only effect 
of the inclusion of (\ref{doubletraceinv}) is the renormalization 
 \be
  c_k \ \rightarrow \ c_k \ + \ 2\,d\, c_{k,l}\;.
 \ee
More generally, the inclusion of any multi-trace invariant  
merely leads to a ($d$-dependent) renormalization of the coefficients $c_k$. 
(We have also verified this at the level of the equations of motion.) Thus, it is sufficient to start with 
the equations~(\ref{offshellfunctions})  
obtained from the single-trace action.

Let us begin with the $\S$ field equation.  
In order to evaluate it efficiently 
we first collect some relations for $\S$ and $\J$ that follow 
from $\dot\S = 2 H \J$  
and $\J^2=-{\bf 1}$: 
\be
\begin{split}
 \S\dot\S^{2k} \ &= \ (-1)^k(2H)^{2k}\S\;, \\
  \dot\S^{2k-1} \ &= \ \dot\S^{2(k-1)}\dot\S 
   \ = \ (-1)^{k-1}(2H)^{2k-1}\J\;. 
\end{split}  
\ee
We can then evaluate the derivative 
 \be
  \frac{d}{dt} \dot\S^{2k-1} \ = \ (2k-1)(-1)^{k-1}(2H)^{2(k-1)} 2\dot{H} \J
  +(-1)^{k-1}(2H)^{2k}\S\;, 
 \ee
and thus compute    
  \be
  \begin{split}
     {d \over dt } \cdS^{2k-1}\ - \dot \Phi \cdS^{2k-1}
    +  \cS  \cdS^{2k} \ = \ &   (-1)^{k-1}(2H)^{2k-2} \big((2k-1)2\dot{H} 
  \, -\dot\Phi (2H) \big)\J \\   
  =\ &  (-1)^{k-1} 2^{2k-1}  \bigl[ (2k-1) \dot H H^{2k-2} 
  - \dot\Phi H^{2k-1} \bigr] \, \J \\    
  =\ &  (-1)^{k-1} 2^{2k-1} \Bigl(\Bigl[ {d\over dt} - \dot\Phi \Bigr]  H^{2k-1} 
  \Bigr) \J  \,. \\   \end{split}
  \ee
Note that here all terms proportional to $\S$ canceled. 
This means that one obtains only a single equation, 
not a matrix equation, as it should be, because we are looking for an equation for the single function $a(t)$.   Writing $E_\S = {\cal E}_\S  \J $ we find
\be\label{ESsimp}
\begin{split}
{\cal E}_\S = & \  - \sum_{k=1}^\infty  (-\alpha')^{k-1}  k\, c_k \,2^{2k} \Bigl[ {d\over dt} - \dot\Phi \Bigr]  H^{2k-1} \,,\\
 = & \  - \Bigl[ {d\over dt} - \dot\Phi \Bigr]\sum_{k=1}^\infty  (-\alpha')^{k-1}  k \, c_k \, 2^{2k}   H^{2k-1} \,.
\end{split}\ee

The last expression  
suggests the definition of a function $f(H)$ of the Hubble parameter $H$.
Including some overall constants for future convenience we write 
 \be\label{fDEFFFF}  
 f(H) \ \equiv \  4d \sum_{k=1}^{\infty}(-\alpha')^{k-1}2^{2k} k\, c_k \,H^{2k-1} \  
  = \ 16\, d\, c_1\, H-128\,d\,\alpha'\,c_2\, H^3 +\cdots \;, 
 \ee
for then the equations (\ref{ESsimp}) can simply be written as 
  \be\label{simpEQQQQ}
   \Big(\frac{{ d}}{{ d}t}-\dot\Phi(t)\Big)f(H) \ = \ 0\;. 
  \ee
This drastic simplification of the equations can be understood as a consequence of the  existence 
of the Noether charges (\ref{NoetherCharge}): for the FLRW ansatz $\Q$  
can be seen to be 
 \be
 \label{Q-noether}
  \Q \ = \ \frac{1}{2d}\,e^{-\Phi(t)} f(H(t))  \begin{pmatrix}  -{\bf 1} & 0 \\[0.7ex] 0 & {\bf 1} \end{pmatrix}  \;. 
 \ee
The conservation law $\dot \Q =0$ 
gives $\frac{{d}}{{d}t}(e^{-\Phi} f(H) ) =0$, which  is  equivalent 
to (\ref{simpEQQQQ}), as anticipated before.    

We now   
 determine the lapse and dilaton equations from (\ref{offshellfunctions}).  
For the lapse equation one obtains 
 \be
 \begin{split}
  \dot\Phi^2 
  \ = \ - 2d \sum_{k=1}^{\infty}(-\alpha')^{k-1}2^{2k}(2k-1)  
   c_k\, H^{2k} 
  \ = \ -\, g(H)\;, 
 \end{split}
 \ee
where we introduced the function 
 \be\label{implicitgDEF}
  g(H) \ \equiv \  2d \sum_{k=1}^{\infty}(-\alpha')^{k-1}2^{2k}(2k-1) c_k\, H^{2k} \ = \ 8\,d\,c_1\, H^2 -\alpha' \,3\cdot 2^5\,d\, c_2\, H^4+\cdots \;. 
 \ee 
The dilaton equation is most conveniently written in the  combination (\ref{finalEQS}) using the lapse equation. One finds 
  \be
  \begin{split}
   0  
   \ = \  \ddot\Phi 
   + 2d \sum_{k=1}^{\infty}(-\alpha')^{k-1}2^{2k}  
   k \,c_k\, H^{2k}\;. 
  \end{split}
  \ee
On the right-hand-side we recognize a multiple of $Hf(H)$. 
Thus, the above equation 
can also be written as 
 \be
  \ddot\Phi + \tfrac{1}{2}H f(H) \ = \ 0\;. 
 \ee 

Summarizing, the three $\alpha'$-completed Friedmann equations take the form 
 \be\label{firstorderEQS} 
  \begin{split}
   \frac{{ d}}{{ d}t}\big(e^{-\Phi} f(H) \big) \ &= \ 0
   \;, \\
  \ddot\Phi + \tfrac{1}{2}H f(H) \ &= \ 0\;, \\
   \dot{\Phi}^2+\, g(H) \ &= \ 0\;, 
  \end{split}
 \ee  
in terms of the functions $f(H)$ and $g(H)$      
defined in (\ref{fDEFFFF}) and (\ref{implicitgDEF}): 
\be
\label{summary-f-g}
\begin{split}
f(H) \ &= \    4d \sum_{k=1}^{\infty}(-\alpha')^{k-1}\, 2^{2k} \ k\, c_k \,H^{2k-1}  \quad \ = \  -2 d\,  H  + {\cal O} (\alpha')  \,,\\  
g(H) \ &= \   2d \sum_{k=1}^{\infty}(-\alpha')^{k-1}2^{2k}(2k\hskip-1pt -1) c_k\, H^{2k} \ = \ - d\, H^2  + \,  {\cal O} (\alpha') \,.    
\end{split} 
\ee
These definitions show that $f$ and $g$  are in fact closely related.  
One readily verifies that 
  \be\label{magicidenitty}
  g'(H) \ = \ H f'(H)\;, 
 \ee
where $'$ denotes differentiation w.r.t.~$H$.

Let us discuss briefly some general properties 
of its solutions. First, as expected,    
they are invariant under the duality 
transformation $a\rightarrow \frac{1}{a}$.
Indeed, under this transformation 
 \be
  H\rightarrow -H\;,\quad  \Phi\rightarrow \Phi\;, \quad f (H)\rightarrow -f(H)\;, \quad g(H)\rightarrow g(H)\;,   
 \ee 
which leaves the equations (\ref{firstorderEQS}) invariant. 
Thus, for any given solution $a(t)$ there is a `dual' solution 
$\tilde{a}(t)\equiv 1/a(t)$.  

Equations~(\ref{firstorderEQS}) are also invariant under time reversal $t\rightarrow -t$. For a given solution  
$a(t)$, $\Phi(t)$, 
there is the time-reversed solution $\tilde{a}(t)\equiv a(-t)$, $\tilde{\Phi}(t)\equiv \Phi(-t)$, with $\tilde H (t) = - H (-t)$.

\subsection{Perturbative solutions}

We now discuss how to solve the $\alpha'$-corrected Friedmann equations (\ref{firstorderEQS}) perturbatively 
in $\alpha'$. To this end it is convenient to introduce the variable $\Omega$ as
an exponential of the dilaton $\Phi$:  
 \be
  \Omega \ \equiv \ e^{-\Phi}\;.
 \ee 
Using the derivatives  
$\dot{\Omega}=-\dot{\Phi}\, \Omega$ and 
$\ddot{\Omega}=(-\ddot{\Phi}+\dot{\Phi}^2)\, \Omega$, 
and taking the difference of the second and third  
equations in (\ref{firstorderEQS}), the set of equations become:  
 \be\label{EQwithOMEGA}
  \begin{split}
    X \ &\equiv \ \frac{{d}}{{d}t}(\Omega f(H)) \ = \ 0\;, \\
   Y \ &\equiv \ \ddot{\Omega} - h(H) \,\Omega \ = \ 0 \;, \\
   Z \ &\equiv \ \dot{\Omega}^2 +  g(H)\,  \Omega^2\ = \ 0\;, 
  \end{split}
 \ee
where, for later use, 
they have been labeled $X,Y, Z$, and we defined
the function $h(H)$: 
 \be
 \begin{split}    
 h(H) \, &\equiv \  \tfrac{1}{2}H f(H)-g(H) \, = \, -2d \sum_{k=2}^{\infty}(-\alpha')^{k-1} 2^{2k} (k-1) c_k H^{2k} 
 \, = \, \alpha'\, 2^5\,d\, c_2\, H^4  +  \cdots \;. 
 \end{split} 
 \ee
Note that $h$, 
in contrast to $f$ and $g$, vanishes to zeroth order in $\alpha'$, 
which will be crucial for our subsequent perturbative construction. 
It will also be crucial to recall that there is one Bianchi identity among the three equations, c.f.~(\ref{LapseBianchi}). 
For the functions defined in (\ref{EQwithOMEGA}) the Bianchi identity  takes the form 
 \be\label{XYZBianchi}
  \frac{{d}Z}{{d}t} \ = \ H\,\Omega\, X \ + \ 2\,\dot{\Omega}\,Y\;. 
 \ee
This is easily verified directly by differentiation recalling   
 that $g'(H) =H f'(H)$.

We will now show how to solve these equations, perturbatively in $\alpha'$, by  
expanding  
 \be\label{PsiHEXPAND}
  \begin{split}
   \Omega(t) \ &= \ \Omega_0(t) + \alpha' \Omega_1(t)+(\alpha')^2\Omega_2(t)+\cdots\;, \\
   H(t) \ &= \ H_0(t) + \alpha' H_1(t) + (\alpha')^2 H_2(t)+\cdots\;. 
  \end{split}
 \ee  
The subscripts in these quantities denote the power of $\alpha'$ that accompanies them in the expansion. 
The first equation, $X=0$, is the Noether conservation, solved once and for all by 
 \be\label{fSOL}
  f(H(t)) \ = \ q\,\Omega^{-1}(t)\;, 
 \ee 
with $q$ a number, proportional to the 
Noether charge $\Q$ (see~(\ref{Q-noether})).   
The question arises whether integration constants such as $q$ have to be treated also as expansions in $\alpha'$ 
by writing $q  =  q_0 + \alpha' q_1+\cdots$. 
This depends on how we 
set out to solve 
the differential equations. 
If we pose initial conditions, say by specifying $\Omega(0)$ and $\dot{\Omega}(0)$, then we have to determine 
all integration constants in terms of these initial data, in which case the value of
$q$, for example,  
does get  $\alpha'$ corrected. 
Here, however, we will follow a different prescription that simplifies the 
analysis.  
Instead of formulating an explicit initial value problem we introduce integration constants whenever 
necessary, but allow ourselves the freedom to set integration constants to zero whose only effect is to `renormalize' constants 
that are already present. As long as these previously introduced integration constants are completely general, this procedure 
does not entail a loss of generality. Then there  
 is no  need for $\alpha'$ expansions of parameters such as~$q$.

We now show how to iteratively solve the above equations to any desired order in $\alpha'$. 
Since $h(H)$ starts at order $\alpha'$, to lowest order $Y=0$ implies 
 \be
  \ddot{\Omega}_0(t) \ = \ 0\qquad \Rightarrow \qquad \Omega_0(t) \ = \  
  \gamma( t-t_0)\;,  
 \ee 
which we solved in terms of two integration constants 
$\gamma$, $t_0$.  
Then, from the expansion of $f(H)$ 
and (\ref{fSOL}), 
 \be
  f(H) \ \equiv \ 16\,d\, c_1 H_0 \ = \ q \, \frac{1}{
  \gamma( t-t_0)} 
   \quad \Rightarrow \quad H_0(t) \ = \ -\frac{q}{2\,d\,
   \gamma}\, \frac{1}{ t-t_0}\;, 
 \ee
where we used the universal $c_1=-\frac{1}{8}$. 
The last equation of (\ref{EQwithOMEGA}), $Z=0$, can now be used to 
to zeroth order in $\alpha'$ to  
establish a relation between $q$ and the other parameters: 
 \be\label{Zlowest}
  0 \ = \ \dot{\Omega}_0^2 +  g(H)  \Omega_0^2 \ = \ 
  \gamma^2-d\,H_0^2(t)
  \gamma^2(t-t_0)^2 \ = \
  \gamma^2-\frac{q^2}{4d}  \;, 
 \ee
and thus 
 \be
  q \ = \ \pm 2\,\sqrt{d}\,|
  \gamma |\;. 
 \ee  
The solutions are then summarized by 
 \be\label{zerothsolution}
  H_0(t) \ = \ \mp\, \frac{{\rm sgn}(
  \gamma)}{\sqrt{d}}\,\frac{1}{t -  t_0} \;, \qquad \Omega_0(t) \ = \  
  \gamma (t -  t_0)\, .    
 \ee   
This  is the complete solution to zeroth order in $\alpha'$  
that is well-defined  for $t>t_0$.   
Before turning to next order, let us express this solution in terms of standard variables. 
From the definition $H(t)=\frac{d}{dt}\ln(a(t))$ we obtain by integration 
 \be
  \ln(a(t)) \ = \ \mp\, \frac{{\rm sgn}(
  \gamma)}{\sqrt{d}}\,\ln(t-t_0)  \ + \ {\rm const.}
 \ee
and thus 
 \be\label{lowestsol1}
  a(t) \ = \ A\, (t-t_0)^{ \mp\, {\rm sgn}(
  \gamma)\frac{1}{\sqrt{d}}}\;, 
 \ee
where $A$ is a new integration constant.  
For the scalar dilaton this yields with (\ref{SFRW}) 
 \be\label{lowestsol2}
  e^{-2\phi} \ = \ (a(t))^{-d} e^{-\Phi} \ = \ (a(t))^{-d}\,\Omega_0 \ \; \propto \ \; \left(t- {t_0}\right)^{1\pm{\rm sgn}(
  \gamma) \sqrt{d}}\;. 
 \ee 
This solution is well-known in the 
literature, see 
eqs.~(4.1)--(4.4) in \cite{Mueller:1989in}, 
which,  for $d=25$ and the special case that all radii are equal, reduce to (\ref{lowestsol1}), (\ref{lowestsol2}) for $\gamma<0$. 
See also \cite{Gasperini:1996fu} for first-order $\alpha'$ corrections. (The relation to our results is more difficult to establish due 
to the ambiguity in field variables and their truncation to four derivatives.)

Let us now turn to the solutions of first order in $\alpha'$. To this order, the equation $Y=0$ implies 
 \be
  \alpha'\, \ddot{\Omega}_{1} \, = \ h(H)\, \Omega_0 \ = \ d(\alpha' \, 2^5 c_2\, H_0^4(t)) \Omega_0 \ = \ 
 \alpha' \,d\, 2^5 c_2\, \frac{1}{d^2}\,\frac{1}{(t -  t_0)^4} \, 
 \gamma\, (t -  t_0)\;, 
 \ee
and so we need to solve 
 \be
  \ddot{\Omega}_{1} \ = \ c_2\,2^5\,\frac{
  \gamma}{{d}}\,\frac{1}{(t -  t_0)^3}\;. 
 \ee
This can be integrated immediately: 
 \be\label{IntegrateOmega1}
  \Omega_1(t) \ = \ c_2\,2^4\,\frac{
  \gamma}{d}\,\frac{1}{t -  t_0} + b_1 t + b_2 \;, 
 \ee 
where $b_1, b_2$ are integration constants. Looking back at the zeroth-order solution (\ref{zerothsolution}) 
we infer that these effectively `renormalize' the previous integration constants 
$\gamma$, $t_0$. Thus, 
without loss of generality we can set $b_1=b_2=0$:  
  \be\label{firstordersolution}
  \Omega_1(t) \ = \ \frac{2^4 c_2
  \gamma}{d}\,\frac{1}{t- t_0}\,.
 \ee 
Next, we have to determine $H_1(t)$.  To this end we use (\ref{fSOL}) and expand to first order in $\alpha'$: 
 \be
  f(H) \ = \ q\, \Omega^{-1} \ = \ q\,\Omega_0^{-1}\big(1+\alpha'\,\Omega_0^{-1}\Omega_1\big)^{-1}
  \ = \ q\,\Big(\Omega_0^{-1} - \alpha'  \, \frac{\Omega_1}{\Omega_0^2}\Big)\;, 
 \ee
which then has to be matched to the definition of $f$ expanded to the same order,  
 \be
  f(H) \ = \ 16\,d\, c_1 H - 128\,d\, \alpha'\, c_2\,H^3 \ = \ 16\,d\,c_1H_0 + 16\,d\,\alpha'\big(c_1 H_1-8\,c_2\, H_0^3)\;. 
 \ee 
Thus, to first order in $\alpha'$ we have to solve 
 \be
   -\frac{q}{16\,d}\, \frac{\Omega_1}{\Omega_0^2} \ = \ c_1 H_1-8\,c_2\, H_0^3\;. 
 \ee 
Using (\ref{zerothsolution}) and (\ref{firstordersolution}) this can be solved for $H_1$, 
 \be
  H_1(t) \ = \ \pm\, \frac{80\,{\rm sgn}(
  \gamma)\, c_2}{d^{3/2}}\,\frac{1}{(t-t_0)^3}\;. 
 \ee

Given the way we have set up the computation, 
the $Z=0$ equation now needs to be satisfied identically, 
for we have no free parameters left to fix by this equation. An explicit computation 
to first order in $\alpha'$ shows that $Z=0$ is indeed satisfied.

\medskip

More generally, it is clear that the iterative procedure of solving the 
equations can be continued to arbitrary orders as follows. 
Consider first the equation $Y=0$,  setting equal the terms 
that are of order $\alpha'^{\,k} $: 
 \be
 \alpha'^{\,k}  \, \ddot{\Omega}_{k} \ = \ (h(H)\,\Omega)_{k}\;. 
 \ee 
Suppose we have solved for $\Omega_0, \ldots, \Omega_{k-1}$, as well as for $H_0, \ldots, H_{k-1}$.  Since $h(H)$ starts at order $\alpha'$, the right-hand side only involves  
$\Omega_i$'s  and $H_i$'s that are already determined. 
Thus $\ddot \Omega_k$ is determined and 
we can directly integrate twice to determine $\Omega_k$. This gives two integration constants as in 
(\ref{IntegrateOmega1}), but as before 
these just renormalize $\gamma, t_0$   
and so we set them to zero. 
With $\Omega_0, \ldots, \Omega_{k}$ now determined we 
next use the equation $X=0$  in the form~(\ref{fSOL})   
 in order to determine $H_{k}$, which completes the next iteration step. 
Upon following the structural dependence on the time-dependent factor $(t-t_0)$ in this computation one may in fact verify 
that the perturbative solutions take the form  
 \be\label{OmegaandH}
  \begin{split}
   H(t) \ &= \ h_0\,\frac{1}{t-t_0} \ + \ \alpha' \,h_1\,\frac{1}{(t-t_0)^3} \ + \ \alpha'^{\,2}\,  h_2\,\frac{1}{(t-t_0)^5}+\cdots  
  \;, \\
   \Omega(t) \ &= \ \omega_0(t-t_0) \  + \ \alpha'\,\omega_1\,\frac{1}{t-t_0} \ + \ \alpha'^{\,2}\,\omega_2\,\frac{1}{(t-t_0)^3}+\cdots 
   \;, 
  \end{split}
 \ee
where $h_n$ and $\omega_n$ are time-independent coefficients, which can in principle be expressed in 
terms of the $c_n$, as done above for the first two. 
Finally, it remains to prove that the $Z=0$ equation is automatically satisfied.
To this end we first note that 
the above expansions of $H$ and $\Omega$ in~(\ref{OmegaandH}) 
determine the time-dependence of $Z= \dot\Omega^2 + g(H) \Omega^2$ to be of the form:   
 \be
  Z(t) \ = \ z_0 \ + \ \alpha' z_1\,\frac{1}{(t-t_0)^2} + \alpha'^{\,2} z_2\,\frac{1}{(t-t_0)^4}+\cdots\;, 
 \ee
where the $z_i$ are time-independent parameters.  Now, since we have satisfied 
$X=Y=0$ in finding~(\ref{OmegaandH}), the Bianchi identity (\ref{XYZBianchi}) implies 
that $Z$ does not depend on time, so actually $z_1=z_2=\ldots=0$.  Therefore, $Z(t)  =  z_0$, 
but since we fixed parameters in (\ref{Zlowest}) so that $z_0=0$ we have established $Z=0$ to all orders in $\alpha'$.

We close this section with a few remarks. The above solutions (\ref{OmegaandH}) seem to suggest that the perturbative expansion breaks down in the high curvature region, since the terms become more 
and more singular as $t\rightarrow t_0$. However, it is conceivable that the entire series converges to a  
regular function (as, for instance, $e^{-\frac{1}{t-t_0}}$ is regular at $t\rightarrow t_0$ although every term in its power 
series diverges). 
Moreover, for late times $t$ the solutions seem to imply that $H$ is driven to zero and thus to Minkowski space, 
but this of course may change once matter contributions are included. In any case, it is important to investigate the 
possible non-perturbative solutions, to which we turn in the next section.\footnote{We thank Robert Brandenberger 
and Jean-Luc Lehners for discussions on these points.}

\section{Non-perturbative solutions}
\label{non-perturb-solu}

In this section we discuss a few aspects of possible solutions that 
are non-perturbative in $\alpha'$. 
In the first subsection we 
solve 
the initial-value formulation for the general equations 
derived above, for generic functions $f(H)$ and $g(H)$. 
In the second subsection 
we give conditions on the functions $f(H)$ and $g(H)$ so that the resulting theory permits de Sitter vacua.

\subsection{Initial-value formulation}

Let us now show how to obtain the general solution given initial
conditions at time $t=0$ of the form:
\be
\Omega(0) \ = \  \omega \,,  \qquad  \dot\Omega (0) \ = \ {1\over \sqrt{\alpha'} }\,  
\gamma
\,.
\ee 
Note that since $\Omega = e^{-\Phi}$ is unit free, $\omega$ and $
\gamma$ are unit-free constants.  
Our strategy will be to solve the first and last equations in (\ref{EQwithOMEGA}), 
\be\label{EQwithOMEGA99}
  \begin{split}
    X \ &\equiv \ \frac{{d}}{{d}t}(\Omega f(H)) \ = \ 0\;, \\
   Z \ &\equiv \ \dot{\Omega}^2 +  g(H)\,  \Omega^2\ = \ 0\; .
  \end{split}
 \ee
The equation $Y=0$ will then hold due to 
the Bianchi identity (\ref{XYZBianchi}), 
 \be\label{XYZBianchi99}
  \frac{{d}Z}{{d}t} \ = \ H\,\Omega\, X \ + \ 2\,\dot{\Omega}\,Y\;, 
 \ee
assuming 
$\dot \Omega \not= 0$ as we will do from now on. 
The case
  $\dot \Omega =0$ will 
  be treated separately and is a bit 
 singular.

 Let us first see that the initial conditions fix the value 
$H(0)$ of the Hubble parameter at $t=0$.
For this consider the second equation in (\ref{EQwithOMEGA99}) which
at $t=0$ gives  
\be\label{Hamiltonianconstr}
\alpha' g (H(0)) \ = \  -  {\gamma^2 \over \omega^2}   \,.
\ee
Note that we could pass to a new function $\tilde g$ defined so that  
\be\label{tildeg}
\alpha' g(H) \ = \  \tilde g (\sqrt{\alpha'} H) 
\ = \ -d(\sqrt{\alpha'}H)^2 \ + \ {\cal O}\big((\sqrt{\alpha'} H)^4\big)\,, 
\ee
which is a unit-free function of the unit-free argument
$\sqrt{\alpha'} H$.    The equation (\ref{Hamiltonianconstr}) then becomes 
\be
\tilde g (\sqrt{\alpha'} H(0)) \ = \ -  {
\gamma^2 \over \omega^2}\,.  
\ee
The possible values of $H(0)$ are the roots of this equation.    In this
way $\Omega(0)$ and $\dot \Omega (0)$ have determined $H(0)$ 
up to a discrete ambiguity. 
Equation  (\ref{tildeg})  makes it  clear  
that we can always solve perturbatively for $H(0)$
but, of course, it could be that non-perturbatively there are no solutions.

Now that we have the initial 
value of $H$ we use the $X=0$ equation
to solve for $\dot \Omega /\Omega$ as a function of $H$:
\be
{\dot \Omega \over \Omega} \ = \ - \, {f'(H)\over f(H)} \dot H\,. 
\ee 
The $Z=0$ equation can be rewritten as 
\be
 {\dot \Omega \over \Omega} = \pm \sqrt{- g(H)} \,.
\ee
The last two equations imply 
\be
 \, {f'(H)\over f(H)} \dot H \ = \ 
 \mp \sqrt{- g(H)}\,, 
\ee
which can be rewritten as 
\be
 \, {f'(H) dH \over f(H)\sqrt{- g(H)}} \ = \  \mp dt\,.
\ee
Integrating from $t=0$ to $t$ then gives
\be
 \int_{H(0)}^{H(t)}    
 \, {f'(H) dH \over f(H)\sqrt{- g(H)}} \ = \  \mp t\,. 
\ee
This is a complete solution.  If the integral is done by finding a $W(H)$ such
that 
\be
dW(H)   = {f'(H) dH \over f(H)\sqrt{- g(H)}} \ = \ -  {2\over H f(H)}  d \sqrt{-g(H)} \,, 
\ee
where the second form follows by $g' = H f'$, c.f.~(\ref{magicidenitty}),  the solution is of the form 
\be
W(H(t)) - W(H(0)) \ = \ \mp t \,. 
\ee
This implicitly determines $H$ as a function of $t$. 
The value of $\Omega(t)$ now follows from the $X=0$ equation in that
\be
\Omega(t) \ = \ {q\over f(H(t))}  \ = \  \omega  {f(H(0))\over f(H(t))} \,,
\ee
where in the second step the constant of the motion $q$ was evaluated 
using the initial condition $\Omega(0) = \omega$ and the value of $H(0)$.

Consider now  
the somewhat singular case of vanishing $\dot \Omega $.  
If $\dot \Omega \equiv 0$ for all $t$ the $Z=0$ equation
implies $g(H)=0$, which fixes  
$H(t)$ to be a constant equal to a zero of $g$. 
We will consider this case in the next subsection. 
  If $t=0$ is the only time for which
$\dot \Omega =0$ then this fixes the initial condition for $H$.

\subsection{Non-perturbative de Sitter solutions}

We now turn to a discussion of possible de Sitter solutions. 
To this end consider the dS metric 
 \be\label{standardDS}
  ds^2 \ = \ -dt^2 + e^{2H_0t}d{\bf x}^2\;, 
 \ee
leading to the scale factor $a(t)=e^{H_0t}$, with 
$H=H_0 \neq 0$  
constant. 
Consider the Friedmann equations  (\ref{firstorderEQS})  copied here for
convenience:
 \be\label{firstorderEQS-vm-bb} 
  \begin{split}
   \frac{{ d}}{{ d}t}\big(e^{-\Phi} f(H) \big) \ &= \ 0
   \;, \\
  \ddot\Phi + \tfrac{1}{2}H f(H) \ &= \ 0\;, \\
   \dot{\Phi}^2+\, g(H) \ &= \ 0\;. 
  \end{split}
 \ee  
With $H$ constant, the first equation  
then implies that ${\Phi}$ is also constant:
\be   
\dot\Phi = 0 \ \ \to \ \ \Phi = \hbox{const.}  
\ee 
Since $H_0 \neq 0$ the second equation implies the vanishing of $f(H_0)$
and the third the vanishing of $g(H_0)$:
\be
\label{sd-condition}
f(H_0) = 0 \,,  \quad  g(H_0) = 0 \,. 
\ee
Note that while the two functions $f(H)$ and $g(H)$ are different, $f(H)$ determines
$g(H)$ via the relations $g'(H)=H f'(H)$ and $g(0)=0$. 
In the two-derivative approximation 
$f(H_0) = - 2d\, H_0$  (see~(\ref{summary-f-g}))  
and therefore the only solution is $H_0=0$, which is not of interest.  

There are, however,   
non-perturbative dS solutions consistent with duality. That is, as we
demonstrate below, there exist
functions $f(H)$ such that dS solutions exist. 
To find a solution to~(\ref{sd-condition}) we integrate the 
relation $g'(H)  =  H f'(H)$ to find:  
 \be\label{gandf}
  g(H) \ = \ Hf(H)-\int_0^H f(H'){\rm d}H'\;, 
 \ee 
where we used 
$g(0)=0$, as required by the definition of $g$.   
To get a vanishing $g(H_0)$ 
when $f(H_0)=0$ we need that the integral of $f(H)$ from zero to $H_0$ vanishes.
This, for example, happens at the non-vanishing zeroes of the sine function.  
More explicitly, we take
 \be\label{fexample}
  f(H) \ \equiv \ -\frac{2d}{\sqrt{\alpha'}}\sin\big(\sqrt{\alpha'}\,H\big) \ = \ -2d
  \sum_{k=1}^{\infty}(-\alpha')^{k-1}\frac{1}{(2k-1)!} H^{2k-1}\;. 
 \ee 
This $f(H) = -2\,d\, H + {\cal O} (\alpha')$ is consistent with the known two-derivative theory.   
From the definition (\ref{fDEFFFF}) it is clear that there are 
coefficients $c_k$ so that 
this holds (in fact $c_k = - {1\over (2k!) 2^{2k}}$).  
With (\ref{gandf}) we then get
 \be
  g(H) \ = \ 
 - {2d\over \alpha'}  \Big(\sqrt{\alpha'}\, H\sin\big(\sqrt{\alpha'}\,H\big) + \cos\big(\sqrt{\alpha'}\,H\big)-1\Big)\;. 
 \ee 
We now see that $f(H_0)=g(H_0)=0$ is satisfied for
 \be
  \sqrt{\alpha'}\,H_0 \, = \, 2\pi n \,,    \quad n \in \mathbb{Z}, \ n\neq 0 \,.
 \ee
This is  
a discrete infinity of dS solutions. 

Although (\ref{fexample}) is surely not the function arising in any string theory, the actual functions  
could 
have the properties required in order to lead to dS vacua in string frame. 
The conditions for this to happen can be stated 
more concisely as follows. Instead of $f(H)$ we consider its integral 
\be  
F(H) \equiv  \int_0^H  f(H') dH'  = 4d\sum_{k=1}^{\infty}(-\alpha')^{k-1}\, 2^{2k-1}\,c_k\, H^{2k}\, ,   
\ee
where the evaluation 
used the expansion~(\ref{summary-f-g}) of
$f(H)$.  
Then, from~(\ref{gandf}), we have that $g(H)$ is given by: 
 \be
  g(H) \ = \ HF'(H) - F(H)\;. 
 \ee
We now infer that if there is an $H_0\neq 0$ so that $F$ and its first derivative vanish 
at this point, 
 \be\label{dScondition}
  F(H_0) \ = \ F'(H_0) \ = \ 0\;, 
 \ee
then $H_0$ is a zero both of $f$ and $g$ and hence 
the theory  permits dS solutions.  
The conditions above require $F$ to attain value zero at maxima or minima.
Such functions
are easily constructed, as we will show below.   
 Since the leading term in $f(H)$ is fixed by the two-derivative theory, we also
have $F(H) = - d \, H^2 + {\cal O} (\alpha')$.

Let us emphasize that the dS solutions discussed above
are non-perturbative in $\alpha'$:   
they are \textit{not} constructed from a solution of the two-derivative equations which is then $\alpha'$-corrected.
 The complete series of all $\alpha'$ corrections 
 needs to be used  
 in order to obtain a solution. 

We demonstrated that the two-derivative theory does not have dS solutions.
The inclusion of the leading $\alpha'$ corrections does not change this.  
But with suitable order $\alpha'^{\,2}$ corrections, dS solutions appear.
This corresponds to  $f(H) \sim  H + \alpha' H^3  + \alpha'^{\,2} H^5$,
and therefore $F(H) \sim H^2 + \alpha' H^4 + \alpha'^{\,2} H^6$.
It is simple to see that this happens for 
\be
F(H) =  - d\,  H^2  ( 1 -   \gamma^2 \alpha' H^2 )^2 \,,
\ee 
where $\gamma$ is an arbitrary constant.  The solution here is
$\sqrt{\alpha'} H_0 = \pm 1/\gamma$.   
Such solutions, however curious, 
cannot be trusted in string theory, which carries an 
infinite number of $\alpha'$ corrections. A reliable solution should either solve the two-derivative equations (and then 
be corrected perturbatively) or be non-perturbative in $\alpha'$.

A  general class of functions $F(H)$ admitting dS solutions is easily obtained.
For a polynomial $w(x)$ the 
condition $w(x_0)=0$  means that $w(x) = (x-x_0) u(x)$, where $u(x)$ is
another polynomial.  The additional condition $w'(x_0)=0$ implies that
$u(x_0)=0$ and thus $w(x) = (x-x_0)^2 v(x)$ for some polynomial $v(x)$.
While $F(H)$ is an infinite series in $H^2$, not a polynomial, the above
remarks show how to produce consistent solutions.  
A general class of $F(H)$ with possible dS Hubble parameters
 $\pm H_0^{(1)}, \ldots \,, \pm  H_0^{(k)}$ takes the form    
\be
F(H) =  - d\,  H^2 \Bigl( 1 + \sum_{p=1}^\infty  d_p \alpha'^{\, p} H^{2p}\Bigr) 
\prod_{i=1}^k \biggl( 1 -  \Big(\frac{H}{H_0^{(i)}} \Big)^2 \biggr)^2 \,.
\ee 
The first factor in parenthesis is an arbitrary series in 
$H^2$ with leading term equal to one.

We close  with a few general remarks on the possible dS solutions. First, they are most likely insufficient as 
phenomenologically viable models, because the natural values of the cosmological constant $\Lambda\sim 1/\alpha'$  
obtained by this mechanism 
are  many orders of magnitude too large. This is the cosmological constant problem. 
Second, they are dS solutions in string frame, while the match with the observable 
dark energy  
is conventionally done in Einstein frame. 
We will now discuss some aspects of the 
Einstein-frame equations.\footnote{We thank   
M.~Gasperini and G. Veneziano 
for explanations on how to relate the two frames.}

Denoting the Einstein frame metric by $G_{\mu\nu}$, 
the required Weyl rescaling reads 
 \be
 \label{scaling-the-metric}
  G_{\mu\nu} \ = \ e^{-\frac{4\phi}{D-2}} g_{\mu\nu}\;,    
  \qquad  D \ = \ d+1  \,,    
 \ee
which implies for the Einstein frame scale factor $a_{E}(t) = e^{-\frac{2}{d-1}\phi(t)} a(t)$.  
Thus, the Einstein frame Hubble parameter reads 
 \be\label{EinsteinHubble}
  H_E \ \equiv  \ {\dot a_E (t) \over a_E(t)}   
   \ =  \ -\frac{2}{d-1}\dot{\phi}+H \ = \ -\frac{1}{d-1}(\dot{\Phi}+H)\;,  
 \ee
 where we used the relation (\ref{phiPhi-der}) between dilaton derivatives:
\be
 \label{phiPhi-der-vmbb}
 2\dot{\phi} \ = \ \dot{\Phi} + d\, H \; .
 \ee 
We next have to recall that the Weyl rescaling (\ref{scaling-the-metric}) implies 
\be
G_{00}  \ = \ \ e^{-\frac{4\phi}{D-2}} g_{00} \ = \  -e^{-\frac{4\phi}{D-2}} \,,
\ee   
since we have set $g_{00} = -n^2= -1$.
To identify a de Sitter solution in the standard form (\ref{standardDS}) we need $G_{00} = -1$, and hence 
we need to reparameterize time.  Let $t'(t)$ be a time such that
$G'_{00}(t') = -1$ and thus 
\be
G_{00}(t) dt^2 \ =\ G'_{00}(t')  dt'^2 \ = \ - dt'^2   
 \quad \to \quad   e^{-\frac{2\phi}{d-1}}  dt \ = \ dt'  \;. 
\ee
This redefinition has no effect on the $g_{ij}$ components of the metric other than 
reparametrizing the time dependence, 
so we have   
\be  
a'_E(t') \ = \  a_E(t(t'))\,.
\ee
Therefore the Hubble parameter of the solution in Einstein frame 
with coordinates $(t',x^i)$ is
\be\label{HEprime}
\begin{split}
H'_E (t') \ &= \ {1\over a'_E(t')} {da'_E (t')\over dt'}    
\ = \ {dt\over dt'} \cdot  {1\over a_E(t)} {da_E (t)\over dt} \ = \ e^{\frac{2\phi}{d-1}}  H_E(t(t')) \\
 \ &= \  - e^{\frac{2\phi}{d-1}}  \frac{1}{d-1} (\dot{\Phi}+H)
\;,
\end{split} 
 \ee
where we used $H_E(t)$ computed in (\ref{EinsteinHubble}).  On the right-hand side dot is a $t$-derivative and all terms are evaluated at $t(t')$.
Since $e^{2\phi} =  (a(t))^d e^\Phi$, we then have 
\be\label{Hprime}
H'_E (t') = - (a(t))^{d\over d-1} \ e^{\Phi\over d-1} \, \frac{1}{d-1} (\dot{\Phi}+H)\;. 
 \ee
Our solutions with $\dot \Phi =0$ and $H= H_0$ would lead to a time-dependent
$H'_E$ through the scale factor $a(t) \sim e^{H_0 t}$.   
Thus, these solutions do not lead to 
dS vacua in Einstein frame.

The general condition for dS vacua in Einstein frame is $\frac{d}{dt'}H_{E}'(t')=0$ and thus, 
by the chain rule, that the right-hand side of (\ref{HEprime}) or  (\ref{Hprime}) is independent of $t$. 
We have not investigated the general problem whether there are solutions with this property, 
but as an illustration for the usefulness of this framework 
we 
present a simple no-go result: 
there are no dS solutions in Einstein frame with constant dilaton $\phi$. 
To this end we observe that $\dot{\phi}=0$ implies with (\ref{phiPhi-der-vmbb}) that $\dot{\Phi}=-d\,H$, 
which with (\ref{HEprime}) yields 
 \be
  H'_{E}(t'(t)) \ = \ e^{\frac{2\phi}{d-1}}H(t)\;. 
 \ee
Since we assume $\phi$ to be constant, $H_{E}'$ is constant if and only if $H$ is constant. 
 With $H$ constant, however, we are back to our previous analysis of dS vacua in string frame, 
 where we concluded that the Friedmann equations (\ref{firstorderEQS-vm-bb}) imply $\dot{\Phi}=0$
 and hence, recalling $\dot{\Phi}=-d\,H$, that $H=0$, leading to 
 $H'_E =0$ 
 and  flat space in Einstein frame.   
 This proves that there are no dS vacua in Einstein frame with constant~$\phi$.

 \bigskip  
 
\section{Conclusions and Outlook}

In this paper we have discussed, to all orders in $\alpha'$, the most general 
duality-invariant 
spacetime actions for a metric, $b$-field and dilaton 
that depend only on time.   
This allowed us to study  
some generic features for cosmological backgrounds 
when all $\alpha'$ corrections are included. In particular, we discussed solutions such as de Sitter vacua that are non-perturbative in $\alpha'$. 
We close with a list of natural follow-up projects and open problems: 

\begin{itemize}

\item The cosmological backgrounds investigated here involved only a single scale factor $a(t)$ in addition to the dilaton. 
In general it should be straightforward to include independent `rolling radii' in different  dimensions, as analyzed to lowest 
order in $\alpha'$ long ago \cite{Mueller:1989in}. Moreover, it would be important to include a $b$-field.  Qualitatively new phenomena may appear in this case. 

\item Eventually it would be crucial to arrive at (semi-)realistic cosmological solutions. 
In particular, one has to make sure that the string coupling given by the dilaton, which here is generally time-dependent, 
does not become too large, for otherwise the classical string theory can no longer be trusted. 
Moreover, matter fields should be included in a duality covariant fashion, as in \cite{Tseytlin:1991xk}.

\item 
An important challenge for any theory of quantum gravity is to show how the big-bang singularity 
could be resolved. One possibility is that 
the higher-derivative $\alpha'$-corrections 
of classical string theory could resolve  the initial singularity, 
leading to a pre-big-bang phase that is smoothly 
connected to the post-big-bang phase \cite{Gasperini:1996fu}. With the framework developed here it should be possible 
to investigate whether this is possible in principle.

\item  In our classification of higher derivative terms, field redefinitions
at each order in $\alpha'$ allowed us to eliminate any terms in which
the field $\S$  had more  than one time derivative or the field $\Phi$ 
had any number of derivatives.  How general is this 
phenomenon in time-dependent field dynamics?   

\item Arguably  a central 
problem is to determine for each string theory the free coefficients defining   
$F(H)$, which determines the functions $f(H)$ and $g(H)$ that appear in the 
$\alpha'$-complete Friedmann equations.  
Perhaps the significantly simplified set-up developed here will suggest a viable strategy. 

\end{itemize}

\subsection*{Acknowledgements} 
We are very grateful to 
Robert Brandenberger,  Maurizio Gasperini, Jean-Luc Lehners, Diego Marques, Krzysztof Meissner, Mark Mueller, 
Ashoke Sen, 
Andrew Tolley and Gabriele Veneziano for useful discussions and correspondence.

The work of O.H.~is supported by the ERC Consolidator Grant ``Symmetries \& Cosmology".

\end{document}